\documentclass[a4paper,preprint,12pt]{elsarticle}
\usepackage[english]{babel}
\usepackage[latin1]{inputenc}
\usepackage{amssymb}
\usepackage{amsmath}         
\usepackage{amsthm}             
\usepackage{pb-diagram}              
\usepackage{epstopdf}      
\usepackage{graphicx}             
 \usepackage{epsfig}       
\usepackage{color}         
\usepackage{fancyhdr} 
 \usepackage{natbib}      
\usepackage{rotating} 
\usepackage[T1]{fontenc}        
\usepackage{url}                
\usepackage{mathtools}                 

\usepackage{vmargin}            
\usepackage[title]{appendix} 

\usepackage{lineno}
\usepackage{setspace}                      
\biboptions{square} 
\usepackage[colorlinks=true,linkcolor= blue,urlcolor=blue,citecolor=blue]{hyperref} 
\usepackage[ruled,vlined]{algorithm2e} 

\newcommand{\hh}{h}

 \newcommand{\bo}[1]{\mathbf{#1}}  
\def\indic{\hbox{1\kern-.24em\hbox{I}}}      

\newcommand{\var}{\mathbb{V}}  
\newcommand{\esp}{\mathbb{E}}

\newcommand{\M}{f}

\newcommand{\h}{\mathbf{h}}
\newcommand{\T}{T}    
\newcommand{\N}{\mathbb{N}}      
\newcommand{\NN}{n}

\newcommand{\norme}[1]{\left|\left| #1 \right|\right|_{2}}

\newcommand{\norminf}[1]{\left|\left| #1 \right|\right|_{\infty}}

\newcommand{\x}{x}
\newcommand{\X}{X}

\newcommand{\R}{\mathbb{R}}

\newtheorem{prop}{Proposition}{\bf}{\it} 
{\bf}{\it}  

\newtheorem{theorem}{Theorem}{\bf}{\it}     
{\bf}{\it}          
\newtheorem{rem}{Remark}{\bf}{\it} 
\newtheorem{corollary}{Corollary}{\bf}{\it} 

\catcode`\"=\active
\catcode`\"=\active
\def"{\og\ignorespaces}
\def"{{\fg}}

\makeatletter
\def\ps@pprintTitle{%
  \let\@oddhead\@empty
  \let\@evenhead\@empty
  \def\@oddfoot{\reset@font\hfil\thepage\hfil}
  \let\@evenfoot\@oddfoot
}
\makeatother

\begin{document}       
\begin{frontmatter}          
\title{Optimal estimators of cross-partial derivatives and surrogates of functions} 
\author[a,b]{Matieyendou Lamboni\footnote{Corresponding author: matieyendou.lamboni[at]gmail.com or [at]univ-guyane.fr; May 30, 2024}}             
\address[a]{University of Guyane, Department DFR-ST, 97346 Cayenne, French Guiana, France}
\address[b]{228-UMR Espace-Dev, University of Guyane, University of R\'eunion, IRD, University of Montpellier, France.}                                             
                                                                   
\begin{abstract}     
 Computing cross-partial derivatives using fewer model runs is relevant in modeling, such as stochastic approximation,  derivative-based ANOVA, exploring complex models, and active subspaces. This paper introduces surrogates of all the cross-partial derivatives of functions by evaluating such functions at $N$ randomized points and using a set of $L$ constraints. Randomized points rely on independent, central, and symmetric variables. The associated estimators, based on $NL$ model runs, reach the optimal rates of convergence (i.e., $\mathcal{O}(N^{-1})$), and the biases of our approximations do not suffer from the curse of dimensionality for a wide class of functions. Such results are used for i) computing the main and upper-bounds of sensitivity indices, and ii) deriving emulators of simulators or surrogates of functions thanks to the derivative-based ANOVA. Simulations are presented to show the accuracy of our emulators and estimators of sensitivity indices. The plug-in estimates of indices using the U-statistics of one sample are numerically much stable. 
\end{abstract}         
\end{frontmatter}                      
%
        
\section{Introduction}        
Derivatives are relevant in modeling, such as inverse problems, first-order and second-order stochastic approximation methods (\cite{robbins51,fabian71,nemirovsky83,polyak90}), exploring complex mathematical models or simulators, derivative-based ANOVA (Db-ANOVA) or exact expansions of functions, and active subspaces.  First-order derivatives or gradients are sometime available in modeling. Instances are:  i) models defined via their rates of change w.r.t. their inputs; ii) implicit functions defined via their derivatives (\cite{cristea17,lamboni23axioms}); iii) cases listed in \cite{morris93,solak02} and the references therein.\\         
  
In FANOVA and sensitivity analysis user and developer communities (see e.g., \cite{hoeffding48a,efron81,sobol93,rabitz99,saltelli00,lamboni22}), screening input variables and interactions of high-dimensional simulators is often performed before building emulators of such models using Gaussian processes, polynomial chaos expansions, SS-ANOVA or other machine learning approaches. Emulators are fast-evaluated models that better approximate complex and/or too expansive simulators. Efficient variance-based screening methods rely on the upper-bounds of generalized sensitivity indices, including Sobol' indices (see \cite{sobol09,kucherenko09,lamboni13,roustant14,lamboni19,lamboni22} for independent inputs and \cite{lamboni21,lamboni23mcap,lamboni24uq,lamboni24} for non-independent variables). Such upper-bounds require the computations of cross-partial derivatives, even for simulators for which these computations are time-demanding or impossible. Also, active subspaces rely on the first-order derivatives for performing dimension reduction and then for approximating complex models (\cite{russi10,constantine14,zahm20}). \\   
            
 For functions with full interactions, all the cross-partial derivatives have been used in the integral representations of the infinitesimal increment of functions (\cite{kubicek15}), and in the unanchored  decompositions of functions in the Sobolev space (\cite{kuo10}). Recently, such derivatives become crucial in the Db-ANOVA representation of every smooth function, such as high-dimensional PDE models. Indeed, it is known in \cite{lamboni22} that every smooth function $\M$ admits an exact Db-ANOVA decomposition, that is, $\forall \, \bo{x} \in \Omega \subseteq \R^d$,    
\begin{equation}    \label{eq:dbanova}         
\displaystyle 
 \M(\bo{x}) = \esp \left[\M(\bo{\X}')\right]  +  \sum_{\substack{v \subseteq \{1, \ldots, d\}\\|v|>0}} \esp_{\bo{\X}'} \left[ \mathcal{D}^{|v|}\M(\bo{\X}') \prod_{k \in v} \frac{G_{k}(\X_{k}') - \indic_{[\X_{k}' \geq x_{k}]}}{g_{k} (\X_{k}') }  \right] \, ,         
\end{equation}  
where $\mathcal{D}^{|v|}\M$ stands for the cross-partial derivative w.r.t. $x_k$ for any $k \in v$; $\bo{\X}' :=\left(\X_1', \ldots, \X_d' \right)$ is a random vector of independent variables, supported on an open $\Omega$ with margins $G_j$'s and densities $g_j$'s (i.e., $\X_j' \sim G_j,\, j=1, \ldots, d$). \\ 
        
Computing all the cross-partial derivatives using a few model runs or evaluations of functions is challenging. Direct computations of accurate cross-partial derivatives have been considered in \cite{bates80,guidotti22} using the generalization of Richardson's extrapolation. Such approximations of all the cross-partial derivatives require a  number of model runs that strongly depends on the dimensionality (i.e., $d$). While adjoint-based methods can provide the  gradients for some POE/PDE-based models using only one simulation of the adjoint models (\cite{dimet86,dimet97,cacuci05,gunzburger03,borzi12,ghanem17}), note that computing the Hessian matrix requires running $d$ second-order adjoint models in general, provided that such models are available (see \cite{dimet86,wang92}). \\         
  
Stochastic perturbations methods or Monte Carlo approaches have been used in stochastic approximations (see e.g., \cite{robbins51,fabian71,nemirovsky83,polyak90}), including derivative-free optimization (see \cite{fabian71,nemirovsky83,polyak90,nemirovsky83,polyak90,agarwal10,bach16,akhavan20,lamboni24axioms} and the references therein), for computing the gradients and Hessian of functions. Such approaches lead to the estimates of gradients using a number of model runs that can be less than the dimensionality  (\cite{patelli10,lamboni24axioms}). While gradients computations and the convergence analysis have been considered in  the first-order stochastic approximations, estimators of the Hessian matrices have been investigated in the second-order stochastic approximations (\cite{fabian71,prashanth16,agarwal17,zhu20,zhu22}). Most of such approaches rely on the Taylor expansions of functions and randomized kernels and/or a random vector that is uniformly distributed on the unit sphere. Nevertheless, independent variables are used in \cite{prashanth16,lamboni24axioms,zhu22}, and the approaches considered in \cite{erdogdu15,zhu22} rely on the Stein identity (\cite{stein04}). Note that the upper-bounds of the biases of such approximations depend on the dimensionality, except in the work \cite{lamboni24axioms} for the gradients only. Moreover, the convergence analysis for more than the second-order cross-partial derivatives are not available according to our knowledge.\\   
                  
Given a smooth function defined on $\R^d$, the motivation of this paper consists in proposing new approaches for deriving surrogates of cross-partial derivatives and derivative-based emulators of functions that        
\begin{itemize}   
\item are simple to be used and generic by making use of $d$ independent variables that are  symmetrically distributed about zero and a set of constraints;     
\item lead to dimension-free upper-bounds of the biases related to the approximations of cross-partial derivatives for a wide class of functions; 
\item provide estimators of cross-partial derivatives that reach the optimal and parametric rates of convergence; 
\item can be used for computing all the cross-partial derivatives and emulators of functions at given points using a few number of model runs.   
\end{itemize}                          
      
In this paper, new expressions of cross-partial derivatives of any order have been derived in Section \ref{sec:dfexpderi} by combining the properties of i) the generalized Richardson extrapolation approach so as to increase the approximations accuracy, and ii) the Monte-Carlo approaches based only on independent random variables that are symmetrically distributed about zero. Such expressions are followed by their order of approximations and biases. We also derive the estimators of such new expressions and  their associated mean squared errors, including the rates of convergence for some classes of functions (see Section \ref{sec:conv}). Section \ref{sec:dfexp} provides the derivative-based emulators of functions, depending on the strength of the interactions or (equivalently) the  cross-partial derivatives thanks to Equation (\ref{eq:dbanova}). The strength of the interactions can be assessed using sensitivity analysis. Thus, Section \ref{sec:sis} deals with the derivation of new expressions of sensitivity indices and their estimators by making use of the proposed surrogates of cross-partial derivatives. Simulations based on test functions are considered  in Section \ref{sec:appli} to show the accuracy of our approach, and we conclude this work in Section~\ref{sec:con}.
                                
\section{Preliminary}    
For an integer $d \in \N\setminus\{0\}$, let $\bo{\X} := (X_1, \, \ldots,\, X_d )$ be a random vector of $d$ independent and continuous variables with marginal cumulative distribution functions (CDFs) $F_j,\, j=1, \ldots, d$ and probability density functions (PDFs) $\rho_j,\, j=1, \ldots, d$.\\
For a non-empty subset $u \subseteq \{1, \ldots, d \}$, we use $|u|$ for its cardinality (i.e., the number of elements in $u$) and $(\sim u) := \{1, \ldots, d \}\setminus u$.  Also, we use $\bo{\X}_u := (\X_j, \, \forall\, j \in u)$ for a subset of inputs, and we have the  partition $\bo{\X} = (\bo{\X}_u, \bo{\X}_{\sim u})$.  Assume that: \\           

\noindent                   
Assumption (A1) : $\bo{X}$ is a random vector of independent variables, supported on $\Omega$; \\
   
Working with partial derivatives requires a specific mathematical space. Given an integer $\NN \in \N\setminus\{0\}$ and an open set $\Omega \subseteq \R^d$, consider a weak partial differentiable function $\M : \Omega \to \R^\NN$ (\cite{zemanian87,strichartz94}) and a subset  $v \subseteq \{1,\, \ldots,\, d\}$ with $|v|>0$. Namely,  we use $\mathcal{D}^{|v|}  \M := \left(\prod_{k \in v} \frac{\partial }{\partial x_{k}} \right) \M$ for the $|v|^{\text{th}}$ weak cross-partial derivatives of each component of $\M$ w.r.t. each $\x_{k}$ with $k \in v$. \\
Likewise, given $\Vec{\boldsymbol{\imath}} :=(i_1, \ldots, i_d)  \in \N^d$, denote    
$\mathcal{D}^{(\Vec{\boldsymbol{\imath}})}\M := \left(\prod_{k=1}^d  \frac{\partial^{i_k} }{\partial x_{k}} \right)\M$ and $\indic_{v}(j) =1$ if $j\in v$ and zero otherwise.  Thus, taking  $\Vec{\boldsymbol{v}} := \left(\indic_{v}(1), \ldots, \indic_{v}(d) \right)$ yields              
$       
\mathcal{D}^{|v|}  \M =  \mathcal{D}^{(\Vec{\boldsymbol{v}})}\M    
$. 
Moreover, denote $(\bo{x})^{\Vec{\boldsymbol{\imath}}} =\bo{x}^{\Vec{\boldsymbol{\imath}}} := \prod_{k=1}^d x_k^{i_k}$, $\Vec{\boldsymbol{\imath}}! := i_1! \ldots i_d!$, and consider the H\"older  space of $\alpha$-smooth functions given by $\forall \, \bo{x}, \bo{y} \in \R^d$   
$$    
\displaystyle            
\mathcal{H}_\alpha := \left\{
\M : \R^d \to \R \, : \,  
\left|\M(\bo{x}) - \sum_{0\leq i_1+\ldots+i_d \leq \alpha-1} 
\frac{\mathcal{D}^{(\Vec{\boldsymbol{\imath}})}\M(\bo{y})}{\Vec{\boldsymbol{\imath}}!} \left(\bo{x}-\bo{y} \right)^{\Vec{\boldsymbol{\imath}}} \right| 
  \leq M_\alpha \norme{\bo{x}- \bo{y}}^\alpha \right\} \, ,     
$$            
with $\alpha\geq 1$, $M_\alpha>0$ and $\mathcal{D}^{(\Vec{\boldsymbol{\imath}})}\M(\bo{y})$ a weak cross-partial derivatives. We use $\norme{\cdot}$ for the Euclidean norm, $||\cdot||_1$ for the $L_1$-norm, $\esp(\cdot)$ for the expectation and $\var(\cdot)$ for the variance.   
  
\section{Surrogates of cross-partial derivatives and new emulators of functions} \label{sec:dfexpderi}
\subsection{New expressions of cross-partial derivatives} \label{sec:dfparder}
This section aims at providing expressions of cross-partial derivatives using the model of interest, and new independent random vectors. We are going to provide approximated expressions of $\mathcal{D}^{|u|}\M(\bo{x})$ for all $u \subseteq \{1, \ldots, d\}$ and the associated orders of approximations.\\        
        
Given $L, \, q \in \N\setminus\{0\}$, consider $\beta_{\ell} \in \R$ with $\ell=1, \ldots, L$, $\boldsymbol{\hh} := (\hh_1, \ldots, \hh_d) \in \R^d_+$, and  denote with $\bo{V} :=(V_1, \ldots, V_d)$ a $d$-dimensional random vectors of independent variables satisfying: $\forall \, j \in \{1, \ldots, d\}$
$$   
 \esp\left[V_j\right] =0; \qquad \esp\left[\left(V_{j} \right)^{2}\right] =\sigma^2;   \qquad \esp\left[\left(V_{j} \right)^{2q+1}\right] =0; \qquad \esp\left[\left(V_{j} \right)^{2q}\right] < +\infty   \, .
$$     
Random vectors of $d$ independent variables that are symmetrically distributed about zero  are instances of $\bo{V}$, including the standard Gaussian random vector and symmetric  uniform distributions about zero.     
         
\noindent    
Denote $\beta_\ell \boldsymbol{\hh}\bo{V}  := (\beta_{\ell}\hh_1 V_1; \ldots, \beta_{\ell} \hh_d V_d)$. The reals $\beta_{\ell}$'s are used for controlling the order of derivatives (i.e., $|u|$) we are interested in, while $V_j$'s help for selecting one particular derivative of order $|u|$. Finally, $h_j$'s aim at defining a neighborhood of a sample point $\bo{x}$ of $\bo{X}$ that is going to be used. Thus, using $\beta_{max} :=\max \left(|\beta_1|, \ldots, |\beta_L|\right)$ and keeping in mind the variance of $\beta_{\ell}\hh_j V_j$, we assume that $\forall \, j \in \{1,\ldots, d\}$\\           
 
Assumption (A2) : $ 
\beta_{max} \hh_j \sigma   \leq 1/2 \quad \mbox{or equivalently} \quad  0< \beta_{max} \hh_j |V_j| \leq 1 \;  \mbox{for bounded} \, V_j's     
$.  \\       
         
Based on the above framework, Theorem \ref{theo:parderord} provides a new expression of the cross-partial derivatives of $\M$. Recall that $|u|$ is the cardinality of $u$ and $\mathcal{D}^{|u|}  \M =  \mathcal{D}^{(\Vec{\boldsymbol{u}})}\M $.   
\begin{theorem} \label{theo:parderord}      
Consider distinct $\beta_\ell$'s, and assume  $\M \in \mathcal{H}_{\alpha}$ with $\alpha \geq |u|+2L$ and (A2) holds. Then, for any $u \subseteq \{1, \ldots, d\}$ with $|u|>0$, there exists  $\alpha_{|u|} \in \{1, \ldots, L\}$ and coefficients $C_{1}^{(|u|)}, \ldots, C_{L}^{(|u|)}$ such that         
\begin{equation}  \label{eq:approxful}        
\displaystyle     
 \mathcal{D}^{|u|}\M(\bo{x}) =  \sum_{\ell=1}^{L} C_{\ell}^{(|u|)} \, \esp \left[ \M\left(\bo{x} + \beta_\ell \boldsymbol{\hh}\bo{V} \right) \prod_{k \in u} \frac{ V_k}{(\hh_k \sigma^2)} \right] +  \mathcal{O}\left( \norme{\bo{\hh}}^{2 \alpha_{|u|}} \right)  \, .                
\end{equation}       
\end{theorem}          
\begin{proof}         
The detailed proof is provided in Appendix \ref{app:theo:parderord}.    
\end{proof}             
                   
In view of Theorem \ref{theo:parderord}, we are able to compute all the cross-partial derivatives using the same evaluations of functions with the same or different order of approximations, depending on the constraints imposed to determine the coefficients $C_{1}^{(|u|)}, \ldots, C_{L}^{(|u|)}$ (see Appendix \ref{app:theo:parderord}).   
While the setting $L=1, \beta_1 =1, \, C_{1}^{(|u|)}=1$ or the constraints 
$  
\sum_{\ell=1}^{L} C_{\ell}^{(|u|)} \beta_{\ell}^{r}= \delta_{|u|,r}; \;  r =0, \ldots, L-2, |u|, \; L-1 \leq |u|$
lead to the order $\mathcal{O}\left( \norme{\bo{\hh}}^{2} \right)$, one can increase that order up to $\mathcal{O}\left( \norme{\bo{\hh}}^{2L} \right)$ by using either 
$        
\sum_{\ell=1}^{L} C_{\ell}^{(|u|)} \beta_{\ell}^{r + |u|}= \delta_{0,r}; \;  r =0, 2, \ldots, 2(L-1)
$ 
or the full constraints given by $    
\sum_{\ell=1}^{L+|u|} C_{\ell}^{(|u|)} \beta_{\ell}^{r}= \delta_{|u|,r}; \;  r =0, \ldots, 2L+|u|-1 $. The last setting is going to improve the approximations and numerical computations of derivatives. Since increasing the number of constraints requires more evaluations of simulators, and in ANOVA-like decomposition of $\M(\bo{\X})$, it is common to neglect the higher-order  components or equivalently the higher-order cross-partial derivatives thanks to Equation (\ref{eq:dbanova}), the following  parsimony number of constraints may be considered.  Given an integer $r^*>0$, controlling the partial derivatives of order up to $r^* \leq L-2$ can be done using the  constraints     
\begin{equation}  \label{eq:consttype1}     
\left\{  
\begin{array}{ll}         
\sum_{\ell=1}^{L} C_{\ell}^{(|u|)} \beta_{\ell}^{r}= \delta_{|u|,r}; \quad  r =0, 1, \ldots, L-1 & 
\mbox{if} \; |u|=1, \ldots, r^* \\   
\sum_{\ell=1}^{L} C_{\ell}^{(|u|)} \beta_{\ell}^{r}= \delta_{|u|,r}; \quad  r =0, \ldots, r^*, |u|, \ldots, |u|+ L-r^*-2  & \mbox{otherwise} \\               
\end{array}       
\right.  \, .                          
\end{equation}     
Equation (\ref{eq:consttype1}) is going to give approximations of all the cross-partial derivatives of $\mathcal{O}\left( \norme{\bo{\hh}}^{\alpha_{|u|}} \right)$ where 
$
o := \left\{
\begin{array}{cl} L-|u| & \, \mbox{if} \, \, 1 \leq |u| \leq r^* \\ 
L-r^*-1  & \,     \mbox{otherwise}  \\         
\end{array} \right.                
$ and $\alpha_{|u|} = o$ if $o$ is even and $o+1$ otherwise. This equation relies on the Vandermonde matrices  and  the generalized Vandermonde matrices, which  ensure the existence and uniqueness of the coefficients for distinct values of $\beta_\ell$'s (i.e., $ \beta_{\ell_1} \neq \beta_{\ell_2} $) because the determinant is of the form $\prod_{1\leq \ell_1 < \ell_2 \leq  L'}\left( \beta_{\ell_1} - \beta_{\ell_2} \right)$ (see \cite{rawashdeh19,ahmed23} for more details and the inverse of such matrices). 
\begin{rem}   \label{rem:betaval} 
 When  $L=1$, we must have $\beta_1=1$, $C_i^{(|u|)} =1,\; \forall\, u \subseteq \{1, \ldots, d\}$. Thus, the coefficient $C_i^{(|u|)}$ does not necessarily depend on $|u|$. Taking $L$ for an even integer, the following nodes may be considered:    
$ 
\left\{\beta_1,\ldots, \beta_{L} \right\} = \left\{\pm 2^{k}, \, k= 0, \ldots, \frac{L-2}{2} \right\}
$.         
When $L$ is odd, one may add $0$ to the above set. Of course,  other possibilities can be considered provided that
$\sum_{\ell=1}^{L} C_{\ell}^{(|u|)} \beta_{\ell}^{|u|}= 1$. 
\end{rem}   
\begin{rem} \label{rem:fo}
For  a given $u_0 \subseteq \{1, \ldots, d\}$,  if we are only interested in all the $|v|^{\mbox{th}}$ cross-partial derivatives with $v \subseteq u_0$, it is better to set $\boldsymbol{\hh}_{\sim u} =\bo{0}$ in Equation (\ref{eq:approxful}).   
 \end{rem}    
    
\begin{rem} Links to other works.\\
Consider $\beta_1=-1, \beta_2=1, \beta_3=0$ and  $V_j \sim \mathcal{N}(0, \, 1)$ with $j=1, \ldots, d$.
Using $L=1$ or $L=2$ or $L=3$, our estimators of the first-order and the second-order cross-partial derivatives are very similar to the results obtained in \cite{zhu22}. \\
   
Using the uniform perturbations and $L=2$ and $L=3$,  our estimators of the first-order and the second-order cross-partial derivatives are similar to those provided in \cite{prashanth16}. However, we are going to see latter that specific uniform distributions allow for obtaining dimension-free upper-bounds of the biases.   
 \end{rem}    
 
\subsection{Upper-bounds of biases}      
To derive precise biases of our approximations provided in Theorem \ref{theo:parderord}, different structural assumptions on the  deterministic functions $\M$ and $\bo{V}$ are going to be considered.  
Assuming $\M \in \mathcal{H}_{\alpha}$ with $\alpha \geq |u|$ is sufficient to define $\mathcal{D}^{|u|}\M(\bo{x})$ for any $u \subseteq \{1, \ldots, d\}$. Note that such an assumption does not depend on the dimensionality $d$.  For the sequel of generality, we are going to provide the upper-bounds of the biases for any value of $L$ by considering two sets of constraints.  \\  

 Denote with $\bo{R} :=(R_1, \ldots, R_d)$ a $d$-dimensional random vector of independent variables that are centered about zero and standardized (i.e., $\esp[R_k^2]=1$, $k=1, \ldots, d$), and $\mathcal{R}_c$ the set of such random vectors. For any $r \in \N$,  define       
$$     
\Gamma_{r} := \sum_{\ell=1}^{L}  \left| C_\ell^{(|u|)} \beta_\ell^{r}   \right|;
\qquad \qquad 
K_{1,L} := \inf_{\bo{R} \in \mathcal{R}_c} \esp\left[ \norme{\bo{R}^2}^{L} \prod_{k \in u} R_k^2 \right] 
 \Gamma_{|u|+2L}  \, .     
$$        
               
\begin{corollary} \label{coro:parderord0}          
Consider distinct $\beta_\ell$'s and the constraints $\sum_{\ell=1}^{L} C_\ell^{(|u|)} \beta_\ell^{|u|+r} = \delta_{0,r}$ with $r=0, 2, 4, \ldots, 2(L-1)$. If $\M \in \mathcal{H}_{|u|+2L}$ and (A2) hold, then   there is $M_{|u|+2L}>0$ such that  
\begin{equation}  \label{eq:bias0}                       
\displaystyle    
  \left|
\mathcal{D}^{|u|}\M(\bo{x}) - \sum_{\ell=1}^{L} C_{\ell}^{(|u|)} \, \esp \left[ \M\left(\bo{x} + \beta_\ell \boldsymbol{\hh}\bo{V} \right) \prod_{k \in u} \frac{ V_k}{\hh_k \sigma^2} \right]
 \right| \leq \sigma^{2L}  M_{|u|+2L} K_{1,L}  \norme{\boldsymbol{\hh}^2}^{L} \, .         
\end{equation}           
Moreover, if $V_k \sim \mathcal{U}(-\xi, \xi)$ with $\xi>0$ and $k=1, \ldots,d$, then
\begin{equation}  \label{eq:bias20}                  
\displaystyle          
   \left| 
\mathcal{D}^{|u|}\M(\bo{x}) - \sum_{\ell=1}^{L} C_{\ell}^{(|u|)} \, \esp \left[ \M\left(\bo{x} + \beta_\ell \boldsymbol{\hh}\bo{V} \right) \prod_{k \in u} \frac{ V_k}{\hh_k \sigma^2} \right] 
 \right| \leq   M_{|u|+2L} \xi^{2L}  ||\boldsymbol{\hh}^2||_1^L  \Gamma_{|u|+2L}   \, .       
\end{equation}          
\end{corollary}                   
\begin{proof}             
See Appendix \ref{app:coro:parderord0} for the detailed proof.        
\end{proof}  
In view of Corollary \ref{coro:parderord0}, one obtains the upper-bounds that do not depend on the dimensionality $d$ by choosing
$ 
\xi =  \sigma =   \frac{1}{\sqrt{d}}, \, \hh_k=\hh     
$ for instance. When $\Gamma_{|u|+2L} >1$, the choice $\xi =  \sigma =   d^{-\frac{1}{2}}  \left(\Gamma_{|u|+2L} \right)^{-\frac{1}{2L}}$ is more appropriate. 
Corollary \ref{coro:parderord0} provides the results for highly smooth functions. To be able to derive the optimal rates of convergence for a wide class of functions (i.e., $\mathcal{H}_{|u|+1}$), Corollary \ref{coro:parderord} starts providing the biases for this class of functions under a specific set of constraints. To that end, define  
$$
K_1 := \inf_{\bo{R} \in \mathcal{R}_c} \esp\left[ \norme{\bo{R}} \prod_{k \in u} R_k^2 \right] \, .
$$       
     
\begin{corollary} \label{coro:parderord}         
For distinct $\beta_\ell$'s, consider $r^* \in \{0, \ldots, |u|-1\}$ and the constraints $\sum_{\ell=1}^{L=r^*+2} C_\ell^{(|u|)} \beta_\ell^{r} = \delta_{|u|,r}$ with $r=0, 1, \ldots, r^*, |u|$. If $\M \in \mathcal{H}_{|u|+1}$ and (A2) hold, then   there is $M_{|u|+1}>0$ such that  
\begin{equation}  \label{eq:bias}                        
\displaystyle                 
  \left|
\mathcal{D}^{|u|}\M(\bo{x}) - \sum_{\ell=1}^{L=2} C_{\ell}^{(|u|)} \, \esp \left[ \M\left(\bo{x} + \beta_\ell \boldsymbol{\hh}\bo{V} \right) \prod_{k \in u} \frac{ V_k}{\hh_k \sigma^2} \right]
 \right| \leq \sigma  M_{|u|+1} K_1  \norme{\boldsymbol{\hh}}  \Gamma_{|u|+1} \, .              
\end{equation}       
Moreover, if $V_k \sim \mathcal{U}(-\xi, \xi)$ with $\xi>0$ and $k=1, \ldots,d$, then
\begin{equation}  \label{eq:bias2}                    
\displaystyle       
   \left|
\mathcal{D}^{|u|}\M(\bo{x}) - \sum_{\ell=1}^{L=2} C_{\ell}^{(|u|)} \, \esp \left[ \M\left(\bo{x} + \beta_\ell \boldsymbol{\hh}\bo{V} \right) \prod_{k \in u} \frac{ V_k}{\hh_k \sigma^2} \right] 
 \right| \leq   \xi  M_{|u|+1} ||\boldsymbol{\hh}||_1  \Gamma_{|u|+1}  \, .   
\end{equation}               
\end{corollary}          
\begin{proof}            
See Appendix \ref{app:coro:parderord}.             
\end{proof}   
Note that the upper-bounds derived in Corollary \ref{coro:parderord} depend on  $r^* \in \{0, \ldots, |u|-1\}$ through  $L=r^*+2$ and $\Gamma_{|u|+1} = \sum_{\ell=1}^{L=r^*+2}  \left| C_\ell^{(|u|)} \beta_\ell^{|u|+1} \right|$. 
Thus, taking $\hh_k=\hh$ and $\xi=1/(d \Gamma_{|u|+1})$ is going to give a dimension-free upper-bound, that does not increase with $r^*$. The crucial role and importance of $r^*$ is going to be highlighted in Section \ref{sec:conv}.   
  
\begin{rem}
When $r^*=0$, which implies that $L=2$, consider $\beta_1 =1, \beta_2 =-1$; $C_{1}^{(|u|)} = 1/2$;  $C_{2}^{(|u|)}= 1/2$ when $|u|$ is even and $C_{2}^{(|u|)}= -1/2$ otherwise. With the above choices, the upper-bounds given by Equations (\ref{eq:bias})-(\ref{eq:bias2}) become, respectively, 
$$
 \sigma  M_{|u|+1} K_1  \norme{\boldsymbol{\hh}} \, ,           
\qquad  \qquad 
\xi   M_{|u|+1} ||\boldsymbol{\hh}||_1  \, .                 
$$
We can check that the same results holds when using $L=1$, $\beta_1=1$ and $C_{1}^{(|u|)} =1$. 
\end{rem}

\begin{rem}    
It is worth noting that we obtain exact approximations of $\mathcal{D}^{|u|}\M(\bo{x})$ in Corollary \ref{coro:parderord0} for the class of functions described by         
 $$  
\mathcal{B}_0 :=\left\{ h \in \mathcal{H}_{\infty} : \; 
\mathcal{D}^{(\Vec{\boldsymbol{\imath}})}h =0, \quad \forall\, \Vec{\boldsymbol{\imath}} \in  \N^d \; \mbox{and} \; ||\Vec{\boldsymbol{\imath}}||_1 \geq |u|+2L     
\right\}\, .    
$$     
In general, exact approximations of $\mathcal{D}^{|u|}\M(\bo{x})$ are obtained when $L\to \infty$  for highly smooth functions.              
\end{rem}
   
\subsection{Convergence analysis} \label{sec:conv}   
Given a sample of $\bo{V}$, that is, 
$\left\{ \bo{V}_i := \left( V_{i, 1}, \ldots, V_{i, d} \right) \right\}_{i=1}^N$ and using Equation (\ref{eq:approxful}),  the method of moments implies that the estimator of $\mathcal{D}^{|u|}\M(\bo{x})$ is given by 
$$    
\displaystyle    
\widehat{\mathcal{D}^{|u|}\M}(\bo{x}) := \frac{1}{N}  \sum_{i=1}^ N \sum_{\ell=1}^{L} C_{\ell}^{(|u|)}  \M\left(\bo{x} + \beta_\ell \boldsymbol{\hh}\bo{V}_i \right) \prod_{k \in u} \frac{ V_{i,k}}{(\hh_k \sigma^2)} \, .     
$$          
         
Statistically, it is common to measure the quality of an estimator using the mean squared error (MSE), including the rates of convergence. The MSEs can also help for determining the optimal value of $\boldsymbol{\hh}$. Theorem \ref{theo:mseall} provides such quantities under different assumptions. To that end, define 
$$  
K_{2, r} := \inf_{\bo{R} \in \mathcal{R}_c} \esp \left[\norme{\bo{R}^2}^{r} \prod_{k \in u} R_{k}^2 \right]  \, .   
$$                
             
\begin{theorem}     \label{theo:mseall}      
For distinct $\beta_\ell$'s, consider $r^* \in \N$ and $\sum_{\ell=1}^{L=r^*+2} C_\ell^{(|u|)} \beta_\ell^{r} = \delta_{|u|,r}$ with $r=0, 1, \ldots, r^*, |u|$ and $r^* \leq |u|-1$.       
 If  $\M \in \mathcal{H}_{|u|+1}$ and (A2) hold, then        
\begin{equation}    \label{eq:mseall}     
\esp\left[ \left( \widehat{\mathcal{D}^{|u|}\M}(\bo{x})- \mathcal{D}^{|u|}\M(\bo{x})\right)^2 \right] \leq  \sigma^2  M_{|u|+1}^2 K_1^2  \Gamma_{|u|+1}^2  \norme{\boldsymbol{\hh}}^2 +  \frac{M_{r^*+1}^2 \Gamma_{r^*+1}^2 K_{2, r^*+1}}{N \sigma^{2(|u|-r^*-1)}\prod_{k \in u} \hh_k^2}   \norme{\boldsymbol{\hh}^2}^{r^*+1}   \, .                    
\end{equation}    
Moreover,  If $V_k \sim \mathcal{U}(-\xi, \xi)$ with $k=1, \ldots, d$, then 
\begin{equation} \label{eq:msealluni}.
\esp\left[ \left( \widehat{\mathcal{D}^{|u|}\M}(\bo{x})- \mathcal{D}^{|u|}\M(\bo{x})\right)^2 \right] \leq   M_{|u|+1}^2 ||\boldsymbol{\hh}||_1^2 \xi^2 \Gamma_{|u|+1}^2  + \frac{3^{|u|}M_{r^*+1}^2 \Gamma_{r^*+1}^2}{N \xi^{2(|u|-r^*-1)}\prod_{k \in u} \hh_k^2}  \norme{\boldsymbol{\hh}}^{2(r^*+1)} \, .                 
\end{equation}              
\end{theorem}                           
\begin{proof}            
See Appendix \ref{app:theo:mseall}  for the detailed proof.           
\end{proof}     	 
Theorem \ref{theo:mseall}  provides the upper-bounds of MSEs for  the anisotropic case. Using a uniform bandwidth, that is, $\hh_k=\hh$ reveals that such  upper-bounds clearly depend on the dimensionality of the function of interest. Indeed,  we can check that the upper-bounds of the MSEs provided in Equations (\ref{eq:mseall})-(\ref{eq:msealluni}) become, respectively,      
$$
  \sigma^2  M_{|u|+1}^2 K_1^2  \Gamma_{|u|+1}^2  d \hh^2 +  \frac{M_{r^*+1}^2 \Gamma_{r^*+1}^2 K_{2, r^*+1}}{N \sigma^{2(|u|-r^*-1)} \hh^{2(|u|-r^*-1)} } d^{\frac{r^*+1}{2}}  \, ,     
$$ 
$$    
\xi^2  M_{|u|+1}^2   \Gamma_{|u|+1}^2 d^2\hh^2 + \frac{3^{|u|}M_{r^*+1}^2 \Gamma_{r^*+1}^2  d^{r^*+1}}{N \xi^{2(|u|-r^*-1)} \hh^{2(|u|-r^*-1)}} \, .          
$$  
By minimizing such upper-bounds with respect to $\hh$, the optimal rates of convergence of the proposed estimators are derived in Corollary \ref{coro.optrate}  
   
\begin{corollary}      \label{coro.optrate}
Under the assumptions made in Theorem  \ref{theo:mseall}, if $r^* < |u|-1$ , then                    
$$
\esp\left[ \left( \widehat{\mathcal{D}^{|u|}\M}(\bo{x})- \mathcal{D}^{|u|}\M(\bo{x})\right)^2 \right] = \mathcal{O}\left(N^{-\frac{1}{|u|-r^*}} d^{1+ \frac{|u|-1}{|u|-r^*}} \right)   \, .                 
$$     
\end{corollary}          
\begin{proof}  
See Appendix \ref{app:coro.optrate} for the detailed proof. 
\end{proof}   
The optimal rates of convergence obtained in Corollary \ref{coro.optrate} are far away from the parametric ones, and such rates decrease with $|u|$. Nevertheless, such optimal rates are function of $d^2$ for any $|u|\geq 2$ using $r^*=1$. The maximum rate of convergence that can be recahed is $N^{-1/2} d^{\frac{|u|+1}{2}}$ by taking $r^*=|u|-2$.  \\  
 To derive the optimal and parametric rates of convergence, let us choose  now $r^*= |u|-1$  and $\hh_k=\hh$ with $k=1, \ldots, d$. Thus, we can see that the second terms of the upper-bounds of the MSEs (provided in Theorem  \ref{theo:mseall}) are function of $d^{|u|}$, but they are independent of $\hh$. This key observation leads to Corollary \ref{coro:optderi1}.    
  
\begin{corollary}      \label{coro:optderi1} 
Under the assumptions made in Theorem  \ref{theo:mseall}, if $r^*= |u|-1$; $\xi =1/(d \Gamma_{|u|+1})$  and $\hh_k=\hh \propto N^{-\gamma/2}$ with $\gamma \in ]1, \, 2[$, then we have        
$$
\esp\left[ \left( \widehat{\mathcal{D}^{|u|}\M}(\bo{x})- \mathcal{D}^{|u|}\M(\bo{x})\right)^2 \right] = \mathcal{O}\left(N^{-1} d^{|u|} \right) \, .          
$$                             
\end{corollary}                   
\begin{proof}     
The proof is straightforward since $\hh^2 \propto N^{-\gamma}$ and $N \hh \to \infty$ if $N \to \infty$.         
\end{proof}                   

It is worth noting that the upper-bound of the squared  bias obtained in Corollary \ref{coro:optderi1} does not depend on the dimensionality thanks to $\xi$. Also, the optimal and parametric rates of convergence are reached by means of $L N=(|u|+1)N$ model evaluations, and such model runs can still be used for computing $\mathcal{D}^{|v|}\M(\bo{x})$ for every $v \subseteq \{1, \ldots, d\}$ with $|v|\leq |u|$. 
Based on the same assumptions, it appears that  the results provided in Corollary \ref{coro:optderi1} are much convenient for $|u|=1, 2$ in higher-dimensions, while those obtained in Corollary \ref{coro.optrate} are well-suited for  higher-dimensions and for higher values of $|u| \in \{1, \ldots, d\}$. \\          

For highly smooth functions and for large values of $|u|$ and $d$, we are able to derive intermediate rates of convergence    
 of the estimator of $\mathcal{D}^{|u|}\M$ (see Theorem \ref{theo:mseaalp}). To that end, consider an interger $L'$, $L= r^*+L'+1$, and denote with $[b]$ the largest integer that is less than $b$ for any real $b$.  
\begin{theorem}     \label{theo:mseaalp}                
For an integer $r^*\leq |u|-2$, consider $\sum_{\ell=1}^{L} C_\ell^{(|u|)} \beta_\ell^{r} = \delta_{|u|,r}$  $r=0, 1, \ldots, r^*, |u|, |u|+2, |u|+4, \ldots, |u|+2(L'-1)$.        
 If  $\M \in \mathcal{H}_{|u|+2L'}$ and (A2) hold, then      
\begin{equation}
\esp\left[ \left( \widehat{\mathcal{D}^{|u|}\M}(\bo{x})- \mathcal{D}^{|u|}\M(\bo{x})\right)^2 \right] \leq  
\sigma^{4L'}  M_{|u|+2L'}^2 K_{1,L'}^2  d^{L'} \hh^{4L'}  +    \frac{M_{r^*+1}^2 \Gamma_{r^*+1}^2 K_{2, r^*+1}}{N \sigma^{2(|u|-r^*-1)} \hh^{2(|u|-r^*-1)} } d^{\frac{r^*+1}{2}} \, .           \nonumber  
\end{equation}     
Moreover, if $V_k \sim \mathcal{U}(-\xi, \xi)$ with $\xi>0$ and $k=1, \ldots,d$, then 
\begin{equation}
\esp\left[ \left( \widehat{\mathcal{D}^{|u|}\M}(\bo{x})- \mathcal{D}^{|u|}\M(\bo{x})\right)^2 \right] \leq 
\xi^{4L'}   M_{|u|+2L'}^2  \Gamma_{|u|+2L'}^2 d^{2L'} \hh^{4L'} + \frac{3^{|u|}M_{r^*+1}^2 \Gamma_{r^*+1}^2  d^{r^*+1}}{N \xi^{2(|u|-r^*-1)} \hh^{2(|u|-r^*-1)}}  \, .     \nonumber 
\end{equation}     
For a given $0\leq \epsilon_{op}< 1$, taking $L'=\left[\frac{(|u|-r^*-1) (1-\epsilon_{op})}{2\epsilon_{op}} \right]$ leads to  
\begin{equation}
\esp\left[ \left( \widehat{\mathcal{D}^{|u|}\M}(\bo{x})- \mathcal{D}^{|u|}\M(\bo{x})\right)^2 \right]  
= \mathcal{O}\left(N^{-1+\epsilon_{op}} d^{|u|(1-\epsilon_{op})}  \right) \, .      
\end{equation}      
\end{theorem}                                 
\begin{proof}          
Detailed proofs are provided in Appendix \ref{app:theo:mseaalp}. 
\end{proof}      	      
            
It comes out that the optimal rate of convergence derived in Theorem \ref{theo:mseaalp} is a trade off between the sample size $N$ and the dimensionality $d$. For instance, when $\epsilon_{op}=1/2$, the optimal rate becomes $\mathcal{O}\left(N^{-1/2} d^{|u|/2)}  \right) $, which improves the rate obtained in Corollary \ref{coro.optrate}, but under different assumptions.    
       
\begin{rem} \label{rem:bo}          
Since the bias vanishes for the class of functions $\mathcal{B}_0$, taking $\sigma =1/\hh$ yiedls 
$
\esp\left[ \left( \widehat{\mathcal{D}^{|u|}\M}(\bo{x})- \mathcal{D}^{|u|}\M(\bo{x})\right)^2 \right] = \mathcal{O}\left(N^{-1}  \right)     
$ (see Appendix \ref{app:rem:bo}).  Note that such an optimal rate of convergence is dimension-free.    
\end{rem}      
    
\subsection{Derivative-based emulators of smooth functions} \label{sec:dfexp}    
Using Equation (\ref{eq:dbanova}) and bearing in mind the  estimators of the cross-derivatives provided in Section \ref{sec:conv}, this section aims at providing surrogates of smooth functions a.k.a. emulators. The general expression of the surrogate of $\M$ is given below.   
\begin{corollary} \label{coro:emu}    
For any $|u| \in \{1, \ldots, d\}$, consider $\sum_{\ell=1}^{L=5} C_\ell^{(|u|)} \beta_\ell^{r} = \delta_{|u|,r}$  for any $r \in \left\{0, 1, \ldots, r^*= \max(|u|-1, 3), |u| \right\}$. Assume that $\M \in \mathcal{H}_{d+1}$ and (A1)-(A2) hold. Then, an approximation of $\M$ at $\bo{x} \in \Omega$ is given by 
\begin{equation}    \label{eq:dbanovaemul}          
\displaystyle 
 \widehat{\M}(\bo{x}) := \esp \left[\M(\bo{\X}')\right]  +  \sum_{\substack{v \subseteq \{1, \ldots, d\}\\|v|>0}} \esp_{\bo{\X}'} \left[ \widehat{\mathcal{D}^{|v|}\M}(\bo{\X}') \prod_{k \in v} \frac{G_{k}(\X_{k}') - \indic_{[\X_{k}' \geq x_{k}]}}{g_{k} (\X_{k}') }  \right] \xrightarrow{P}  \M(\bo{x}) \, .         
\end{equation}                           
\end{corollary}                  
The above plug-in estimator is consistent using the law of large numbers. For a given $\bo{x}$, it is worth noting that the choice of $G_j$'s is arbitrary, provided that such distributions are supported on an open neighborhood of $\bo{x}$.   \\   
 Often, the higher-order cross-partial derivatives or equivalently the  higher-order interactions among the model inputs almost vanish, leading to consider the truncated expressions. Given an integer $s$ with $0<s \ll d$ and keeping in mind the ANOVA decomposition, consider the class of functions that admit at most the $s^{\mbox{th}}$-order interactions, that is,  
$
\mathcal{A}_s :=\left\{h :\R^d \to R :\, h(\bo{x}) = \sum_{\substack{v \subseteq \{1, \ldots, d\}\\ |v|\leq s}} h_v(\bo{x}_v)  \right\}          
$.  
Truncating the functional expansion is a standard practice within the ANOVA-community, that is, $ s \ll d$ is assumed in higher-dimensions (\cite{efron81,rabitz99}). For such a class of functions, requiring $\M \in \mathcal{H}_{\alpha_s}$ with $\alpha_s \geq s$ is sufficient to derive our results. Thus, the truncated surrogate of $\M$ is given by    
$$
\displaystyle 
 \widehat{\M_s}(\bo{x}) := \esp \left[\M(\bo{\X}')\right]  +  \sum_{\substack{v \subseteq \{1, \ldots, d\}\\0< |v| \leq s}} \esp_{\bo{\X}'} \left[ \widehat{\mathcal{D}^{|v|}\M}(\bo{\X}') \prod_{k \in v} \frac{G_{k}(\X_{k}') - \indic_{[\X_{k}' \geq x_{k}]}}{g_{k} (\X_{k}') }  \right] \, .     
$$ 
       
Under the assumptions made in Corollary \ref{coro:emu},  $\widehat{\M_s}(\bo{x})$ reaches the optimal and parametric rate of convergence for the class of functions $\mathcal{A}_3$. For instance, taking $s=1$ leads to the first-order emulator of $\M$, which relies only on the gradient information. Thus, $\widehat{\M_{s=1}}$ provides accurate estimates of additive models of the form $\sum_{j=1}^d h_j(x_j)$, where $h_j$'s are given functions. Likewise, $\widehat{\M_{2}}$ allows for incorporating the second-order terms, but it requires the second-order cross-partial derivatives. Thus, it is relevant to find the class of functions $\mathcal{A}_s$ which contains the model of interest before building emulators. The following section deals with such issues.      
			          
\section{Applications: computing sensitivity indices} \label{sec:sis}           
In high-dimensional settings, reducing the dimension of functions is often done by using screening measures, that is, measures that can be used for quickly identifying non-relevant input variables.  Screening measures based on the upper-bounds of the total sensitivity indices rely on derivatives (\cite{morris91,kucherenko09,lamboni13,roustant14,roustant17,lamboni22}). This section aims at providing optimal computations of upper-bounds of the total indices, followed by the computations of the main  indices using derivatives.\\            
 
By evaluating the function $\M$ given by Equation (\ref{eq:dbanova}) at a random vector $\bo{\X}$ using $G_j=F_j$, one obtains a random vector of the model outputs. Generalized sensitivity indices, including Sobol' indices rely on the variance-covariance of sensitivity functionals (SFs),  which are also random vectors containing the information about the overall contributions of inputs (\cite{lamboni16,lamboni18a,lamboni18,lamboni19,lamboni22}). The derivative-based expressions of SFs are given below (see \cite{lamboni22} for more details). Given $u \subseteq \{1, \ldots, d\}$, the interaction SF of the inputs $\bo{\X}_u$ is given by 
\begin{equation} \label{eq:sfin0}         
\displaystyle        
\M_u(\bo{\X}_u)  :=  
\esp_{\bo{\X}'} \left[ \mathcal{D}^{|u|}\M(\bo{\X}') \prod_{k \in u} \frac{F_{k}(\X_{k}') - \indic_{[\X_{k}' \geq \X_{k}]}}{\rho_{k} (\X_{k}') } \right]   \, , \nonumber 
\end{equation}
and the first-order SF of $\bo{\X}_u$ is given by  
$$     
\displaystyle                
\M_u^{fo}(\bo{\X}_u) := \sum_{\substack{v, v \subseteq u \\|v|>0}} \M_v(\bo{\X}_v)  \, .
$$   
Likewise, the total-interaction SF of $\bo{\X}_u$ is given by (\cite{lamboni22})
\begin{equation} \label{eq:sfti0}
\displaystyle     
\M_u^{sup}(\bo{\X}) := \sum_{\substack{v\subseteq \{1, \ldots, d\}\\ u \subseteq v }} \M_v(\bo{\X}_v) =  \esp_{\bo{\X}'} \left[ 
\mathcal{D}^{|u|} \M\left(\bo{\X}'_{u},\bo{\X}_{\sim u}\right) 
 \prod_{k \in u} \frac{F_{k}(\X_{k}') - \indic_{[\X_{k}' \geq \X_{k}]}}{\rho_{k} (\X_{k}') }  \right]   \, ,       \nonumber  
\end{equation} 
and the total SF of $\bo{\X}_u$ is given as (\cite{lamboni22})    
$$   
\displaystyle         
\M_u^{tot}(\bo{\X}) :=  \sum_{\substack{v\subseteq \{1, \ldots, d\}\\ u \cap v \neq \emptyset}} \M_v(\bo{\X}_v)  =  
\sum_{\substack{v, v \subseteq u \\|v|>0}}
\esp_{\bo{\X}'} \left[ \mathcal{D}^{|v|}\M(\bo{\X}'_u, \bo{\X}_{\sim u}) \prod_{k \in v} \frac{F_{k}(\X_{k}') - \indic_{[\X_{k}' \geq \X_{k}]}}{\rho_{k} (\X_{k}') } \right]   \, . \nonumber     
$$       
 
For a single input $\X_j$, we have $\M_j(\X_j)=\M_j^{fo}(\X_j)$ and $\M_j^{tot}(\bo{\X})= \M_j^{sup}(\bo{\X})$. Among similarity measures (\cite{lamboni24uq,lamboni24}), taking the variance-covariances of SFs, that is, $\Sigma_u := \var\left[\M_u(\bo{\X}_u) \right]$, $\Sigma^{sup}_u := \var\left[ \M_u^{sup}(\bo{\X}) \right]$ and $\Sigma^{tot}_u:= \var\left[\M_u^{tot}(\bo{\X}) \right]$, leads to (\cite{lamboni22})     
\begin{equation} \label{eq:intsigu}   
\displaystyle   
  \Sigma_u =  \esp \left[ \mathcal{D}^{|u|}\M(\bo{\X}) \mathcal{D}^{|u|}\M(\bo{\X}') 	\prod_{k \in u}       
  \frac{F_{k}\left(\min\left(\X_{k},\, \X_{k}'\right) \right) -F_{k}(\X_{k})F_{k}(\X_{k}')}{\rho_{k} (\X_{k}) \rho_{k} (\X_{k}')}  \right] \, ;          
\end{equation}     
\begin{equation} \label{eq:linhint}  
\displaystyle   
\Sigma_u  \leq \Sigma^{sup}_u  \leq  \Sigma^{ub}_u := \frac{1}{2^{|u|}}
 \esp \left[ \left( \mathcal{D}^{|u|}\M(\bo{\X}) \right)^2 
	\prod_{k \in u}   
  \frac{F_{k}\left(\X_{k}\right)\left[1 -F_{k}(\X_{k})\right]}{\left(\rho_{k} (\X_{k})\right)^2}  \right] \, .  
\end{equation}     
Thus, $\Sigma^{ub}_u$ is the upper-bound of $\Sigma^{sup}_u$. Likewise, $\Sigma^{ub}_j$ is the upper-bound of $\Sigma^{tot}_j$ (i.e., $\Sigma^{tot}_j \leq \Sigma^{ub}_j$), and it can be used for screening the input variables. \\   
To provide new expressions of the screening measures and the main induces in the following proposition, denote with $\bo{V}'$ an i.i.d. copy of $\bo{V}$, and assume that \\

Assumption (A3) : $\M(\bo{\X})$ has finite second-order moments. \\                  
\begin{prop}   \label{prop:totintcov}      
Under the assumptions made in Corollary  \ref{coro:optderi1}, assume (A1)-(A3) hold. Then,        
\begin{eqnarray}  \label{eq:sigudf}      
\Sigma_u  &=&  \sum_{\substack{\ell_1=1  \\ \ell_2=1}}^{L}  C_{\ell_1}^{(|u|)}   C_{\ell_2}^{(|u|)}   \esp \left[ \M\left(\bo{\X} + \beta_{\ell_1} \boldsymbol{\hh}\bo{V} \right)  
\M\left(\bo{\X}' + \beta_{\ell_2} \boldsymbol{\hh}\bo{V}' \right)
 \right. \nonumber \\     
& & \left.    \times 
\prod_{k \in u} \frac{  V_k  V_k'}{(\hh_k^2 \sigma^4 )} \left(  
  \frac{F_{k}\left(\min\left(\X_{k},\, \X_{k}'\right) \right) -F_{k}(\X_{k})F_{k}(\X_{k}')}{\rho_{k} (\X_{k}) \rho_{k} (\X_{k}')} \right) \right] + \mathcal{O}\left( \norme{\bo{\h}}^{2} \right) \, ;         
\end{eqnarray}            
\begin{eqnarray}  \label{eq:ubsigudf}      
\Sigma^{ub}_u  &=& 
\frac{1}{2^{|u|}}
  \sum_{\substack{\ell_1=1  \\ \ell_2=1}}^{L}  C_{\ell_1}^{(|u|)}   C_{\ell_2}^{(|u|)}   \esp \left[ \M\left(\bo{\X} + \beta_{\ell_1} \boldsymbol{\hh}\bo{V} \right)  
\M\left(\bo{\X} + \beta_{\ell_2} \boldsymbol{\hh}\bo{V}' \right)  
 \right. \nonumber \\     
& & \left.  \times 
\prod_{k \in u} \frac{  V_k  V_k'}{\hh_k^2 \sigma^4} \left(  
  \frac{F_{k}\left(\X_{k}\right)\left[1 -F_{k}(\X_{k})\right]}{\left(\rho_{k} (\X_{k})\right)^2} \right) \right] + \mathcal{O}\left( \norme{\bo{\h}}^{2} \right) \, .        
\end{eqnarray}                
\end{prop}                   
\begin{proof}
Bearing in mind Equations (\ref{eq:approxful}), (\ref{eq:intsigu}) and  (\ref{eq:linhint}), the proof of Proposition \ref{prop:totintcov} is straightforward.      
\end{proof}  
  
The method of moments allows for deriving the estimators of $\Sigma_u$ and $\Sigma^{ub}_u$ for all $u \subseteq \{1, \ldots, d\}$. For screening inputs of models, we are going to provide the estimators of $\Sigma_j$ and $\Sigma^{ub}_j$ for any $j \in \{1, \ldots, d\}$.  To that end, we are given four independent samples, that is,  $\left\{\bo{\X}_i \right\}_{i=1}^N$ from $\bo{\X}$,  $\left\{\bo{\X}_i' \right\}_{i=1}^N$ from $\bo{\X}'$,  $\left\{\bo{V}_i \right\}_{i=1}^N$ from $\bo{V}$ and $\left\{\bo{V}'_i \right\}_{i=1}^N$ from $\bo{V}'$. Consistent estimators of $\Sigma_j$ and  $\Sigma^{ub}_j$ are respectively given by  
\begin{eqnarray}  \label{eq:sigudfest}      
\widehat{\Sigma_j}  & := & \sum_{i=1}^N \sum_{\substack{\ell_1=1  \\ \ell_2=1}}^{L} \frac{C_{\ell_1}^{(1)}   C_{\ell_2}^{(1)} }{N} \M\left(\bo{\X}_i + \beta_{\ell_1} \boldsymbol{\hh}\bo{V}_i \right)  
\M\left(\bo{\X}'_i + \beta_{\ell_2} \boldsymbol{\hh}\bo{V}'_i \right)
\frac{  V_{i,j}  V_{i,j}'}{\hh_j^2 \sigma^4} \nonumber \\
& & \left(  
  \frac{F_{j}\left(\min\left(\X_{i,j},\, \X_{i,j}'\right) \right) -F_{j}(\X_{i,j})F_{j}(\X_{i,j}')}{\rho_{j} (\X_{i,j}) \rho_{j} (\X_{i,j}')} \right)  \, ;    
\end{eqnarray}            
\begin{equation}  \label{eq:sigudfestub}     
\displaystyle  
\widehat{\Sigma^{ub}_j}  :=    \sum_{i=1}^N  
  \sum_{\substack{\ell_1=1  \\ \ell_2=1}}^{L} 
	\frac{C_{\ell_1}^{(1)}   C_{\ell_2}^{(1)} }{2 N}
 \M\left(\bo{\X}_i + \beta_{\ell_1} \boldsymbol{\hh}\bo{V}_i \right)  
\M\left(\bo{\X}_i + \beta_{\ell_2} \boldsymbol{\hh}\bo{V}'_i \right) 
\frac{  V_{i,j}  V_{i,j}'}{\hh_j^2 \sigma^4} \left(  
  \frac{F_{j}\left(\X_{i,j}\right)\left[1 -F_{j}(\X_{i,j})\right]}{\left(\rho_{j} (\X_{i,j})\right)^2} \right) \, .    
\end{equation}        

The above (direct) estimators require $3 L N$ model runs for obtaining the estimates for any $j \in \{1, \ldots, d\}$. Additionally to such estimators, we are going to derive the plug-in estimators, which are relevant in the presence of a given data about the estimates of the first-order derivatives. To provide such estimators, denote with $\left\{  \widehat{\mathcal{D}^{|\{j\}|}\M}(\bo{\X}_i) \right\}_{i=1}^{N_1}$ a sample of $N_1>1$ known or estimates of the first-order derivatives (i.e., $|u|=1$).  Using Equation (\ref{eq:approxful}), such estimates are obtained by considering $L=1$ or $L=2$ or $L=3$ and $N$.  Keeping in mind Equations (\ref{eq:intsigu})-(\ref{eq:linhint}) and the U-statistic theory for one sample, the plug-in estimator of the main index of $\X_{j}$ is given by 
\begin{equation} \label{eq:intsiguest}    
\displaystyle   
 \widehat{\Sigma_j}' := \frac{2}{N_1(N_1-1)} \sum_{1\leq i_1 < i_2 \leq N_1}  
 \widehat{\mathcal{D}^{|\{j\}|}\M}(\bo{\X}_{i_1})
 \widehat{\mathcal{D}^{|\{j\}|}\M}(\bo{\X}_{i_2})      
  \frac{F_{j}\left(\min\left(\X_{i_1, j},\, \X_{i_2, j}\right) \right) -F_{j}(\X_{i_1,j})F_{j}(\X_{i_2,j})}{\rho_{j} (\X_{i_1,j}) \rho_{j} (\X_{i_2,j})}  \, .          
\end{equation}         
Likewise, the plug-in estimator of the upper-bound of the total index of $\X_{j}$ is given by 
\begin{equation} \label{eq:intsiguestub}   
\displaystyle   
 \widehat{\Sigma_j^{ub}}'  :=   \frac{1}{2N_1} \sum_{i=1}^{N_1}
\left[ \widehat{\mathcal{D}^{|\{j\}|}\M}(\bo{\X}_{i}) \right]^2 
  \frac{F_{j}\left(\X_{i,j}\right)\left[1 -F_{j}(\X_{i,j})\right]}{\left[\rho_{j} (\X_{i,j})\right]^2}  \, .            
\end{equation}       
Note that the  plug-in estimators are consistent and require a total of $LN_1 N_0$ model runs for computing such indices, where $N_0$ is the number of model runs used for computing the gradient of $\M$ at $\bo{\X}_i$. 
  
\section{Illustrations: screening and emulators of models} \label{sec:appli} 
\subsection{Test functions} \label{sec:testfun} 
\subsubsection{Ishigami's function ($d=3$)} 
The Ishigami function includes three independent inputs following a uniform distribution on $[-\pi,\, \pi]$, and it is given by 
\begin{equation} \label{eq:ishi}
f(\bo{\x}) =  \sin(\x_1)+ 7\sin^2(\x_2) + 0.1 \x_3^4 \sin(\x_1) \, .  \nonumber 
\end{equation}
The sensitivity indices are  $S_{1} = 0.3139$, $S_{2} = 0.4424$, $S_{3} = 0.0$,
$S_{T_1} = 0.567$, $S_{T_2} = 0.442$, and $S_{T_3} = 0.243$. 
   
\subsubsection{Sobol's g-function ($d=10$)}
The g-function (\cite{homma96}) includes ten independent inputs following a uniform distribution on $[0,\, 1]$, and it is defined as follows:  
\begin{equation} \label{eq:gsobol}
f(\bo{\x}) = \prod_{j=1}^{d=10}\frac{ |4 \x_j - 2| + a_j}{1 + a_j}\, .     \nonumber 
\end{equation}  
Note that such a function is differentiable almost everywhere. According to the values of $\bo{a} = (a_j, \, j = 1, \, 2, \, \ldots,\, d)$, this function has different properties (\cite{kucherenko09}):
\begin{itemize}
\item if $ \bo{a} = [0,\, 0,\, 6.52,\, 6.52,\, 6.52,\, 6.52,\, 6.52,\, 6.52,\, 6.52,\, 6.52]^\T$, the values of sensitivity indices are $S_1 = S_2 = 0.39$, $S_j =0.0069$,   $\forall \, j > 2 $, $S_{T_1} = S_{T_2} = 0.54$,  and $S_{T_j} = 0.013$,  $\forall \,  j > 2 $. Thus, this function has a low effective dimension (function of type A), and it belongs to $\mathcal{A}_s$ with $s>1$ (see Section \ref{sec:dfexp});  
\item  if $\bo{a} = [50,\, 50,\, 50,\, 50,\, 50,\, 50,\, 50,\, 50,\, 50,\, 50]^\T$, the first and total indices are given as follows: $S_j = S_{T_j} = 0.1$,  $\forall \, j \in  \{ 1,\, 2,\, \ldots,\, d\}$. Thus, all inputs are important, but there is no interaction among these inputs. This function has a high effective dimension (function of type B). Note that it belongs to $\mathcal{A}_1$;    
\item if $\bo{a} = [0,\, 0,\, 0,\, 0,\, 0,\, 0,\, 0,\, 0,\, 0,\, 0]^\T$, the function belongs to the class of functions with important interactions among inputs. Indeed, we have $S_j = 0.02$ and  $ S_{T_j} = 0.27$,   $\forall\, j \in  \{ 1,\, 2,\, \ldots,\, d\}$. All the inputs are relevant due to important interactions (function of type C). Then, this function belongs to $\mathcal{A}_s$ with $s\geq 2$.    
 \end{itemize}                                        
										       
\subsection{Numerical comparisons of estimators} \label{sec:compde} \label{sec:indices}  
This section provides a comparison of the direct and plug-in estimators of the main indices and the upper-bounds of the total indices using the test functions of Section \ref{sec:testfun}. Different total budgets for the model evaluations are considered in this paper, that is, $N=500, 1000, 1500, 2000, 3000, 5000, 10000, 15000, 20000$. In the case of the plug-in estimators, we used $N_0=2d$. For generating different random values, Sobol's sequence (scrambled =3) from the R-package randtoolbox (\cite{dutang13}) is used. We replicated each estimation $30$ times by randomly choosing the seed, and the reported results are the average of the $30$ estimates.\\     
   
\begin{figure}[!hbp]  
\begin{center}
\includegraphics[height=15cm,width=9cm,angle=270]{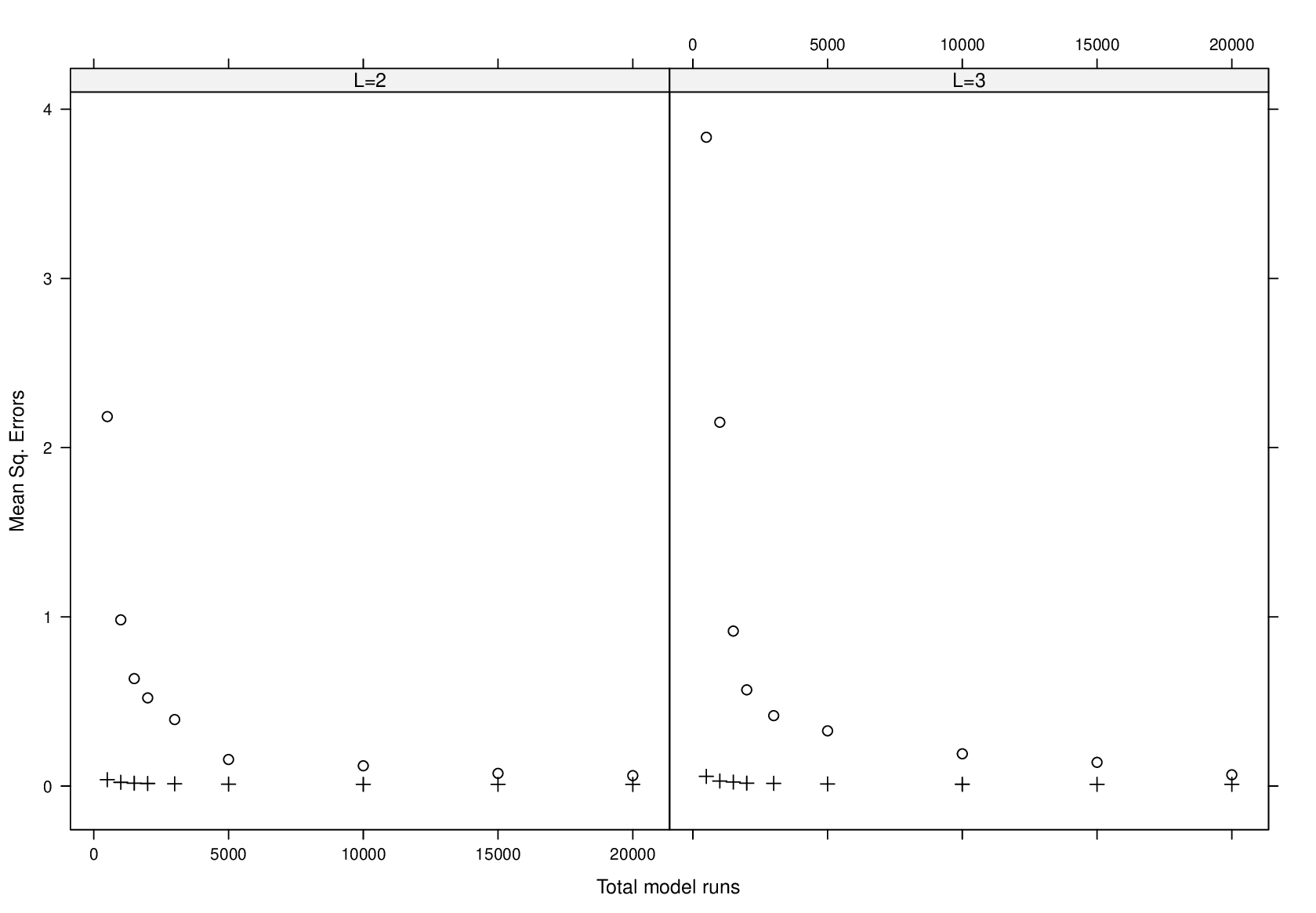}      
\end{center} 
\caption{Average of $d=3$ mean squared errors using the Ishigami function ($\circ$ the direct estimator (\ref{eq:sigudfest})   and $+$ for the plug-in estimator (\ref{eq:intsiguest})).}         
 \label{fig:ish}           
\end{figure}

\begin{figure}[!hbp]         
\begin{center}
\includegraphics[height=15cm,width=9cm,angle=270]{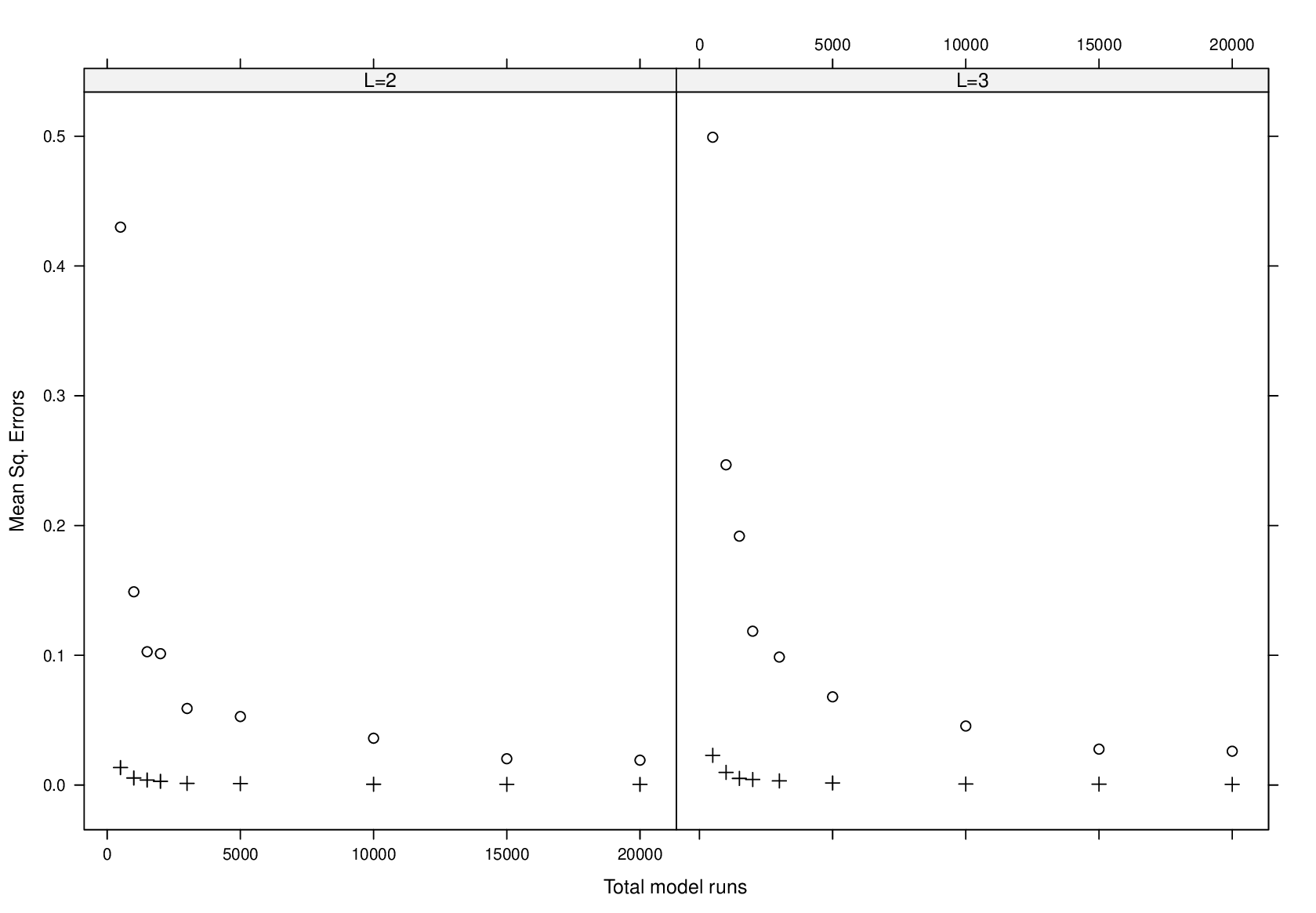}    
\end{center} 
\caption{Average of $d=10$ mean squared errors using the g-function of type A ($\circ$ the direct estimator (\ref{eq:sigudfest})   and $+$ for the plug-in estimator (\ref{eq:intsiguest})).}      
 \label{fig:sobA} 
\end{figure}

\begin{figure}[!hbp]         
\begin{center}
\includegraphics[height=15cm,width=9cm,angle=270]{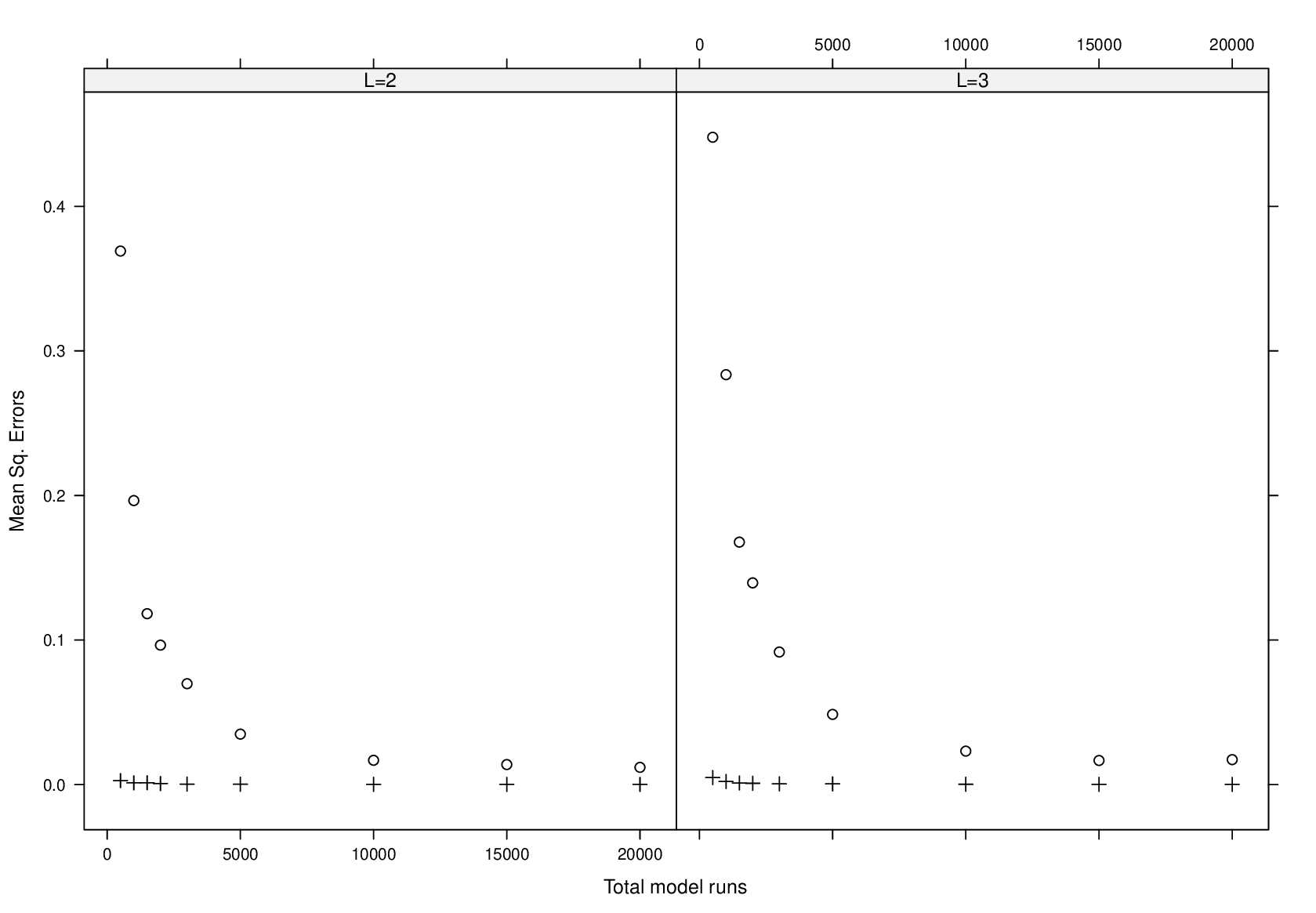}    
\end{center} 
\caption{Average of $d=10$ mean squared errors using the g-function of type B ($\circ$ the direct estimator (\ref{eq:sigudfest})   and $+$ for the plug-in estimator (\ref{eq:intsiguest})).}      
 \label{fig:sobB}          
\end{figure}

\begin{figure}[!hbp]           
\begin{center}
\includegraphics[height=15cm,width=9cm,angle=270]{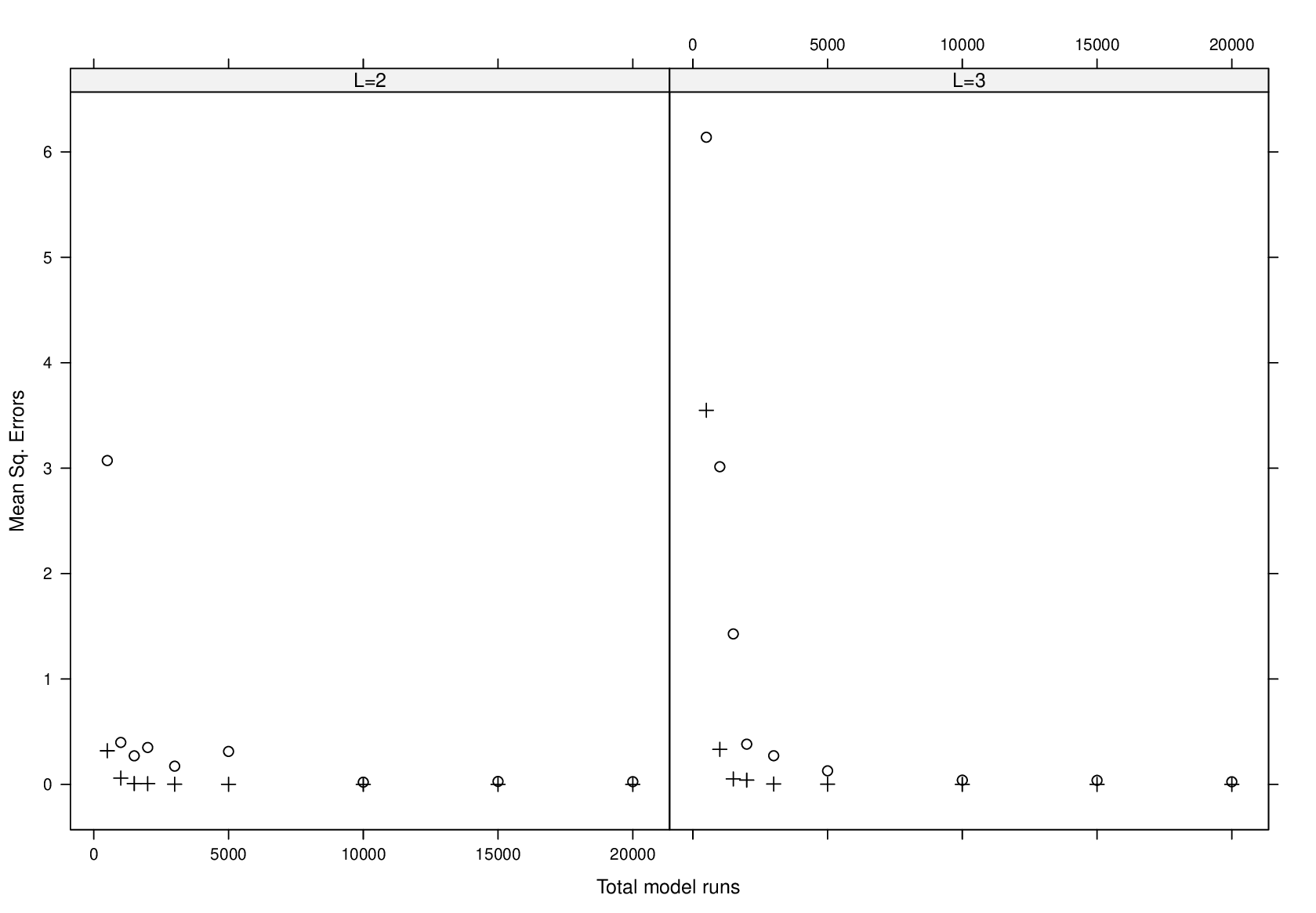}    
\end{center} 
\caption{Average of $d=10$ mean squared errors using the g-function of type C ($\circ$ the direct estimator (\ref{eq:sigudfest})   and $+$ for the plug-in estimator (\ref{eq:intsiguest})).}         
 \label{fig:sobC}                  
\end{figure}

 Figures \ref{fig:ish}-\ref{fig:sobC} show the mean squared errors related to the estimates of the main indices for the Ishigami function, the g-functions of type A, type B and type C, respectively. Each figure depicts the results for $L=2$ and $L=3$.  All the figures show the convergence and accuracy of our estimates using either $L=2$ or $L=3$.  
 
\begin{figure}[!hbp]  
\begin{center}
\includegraphics[height=15cm,width=9cm,angle=270]{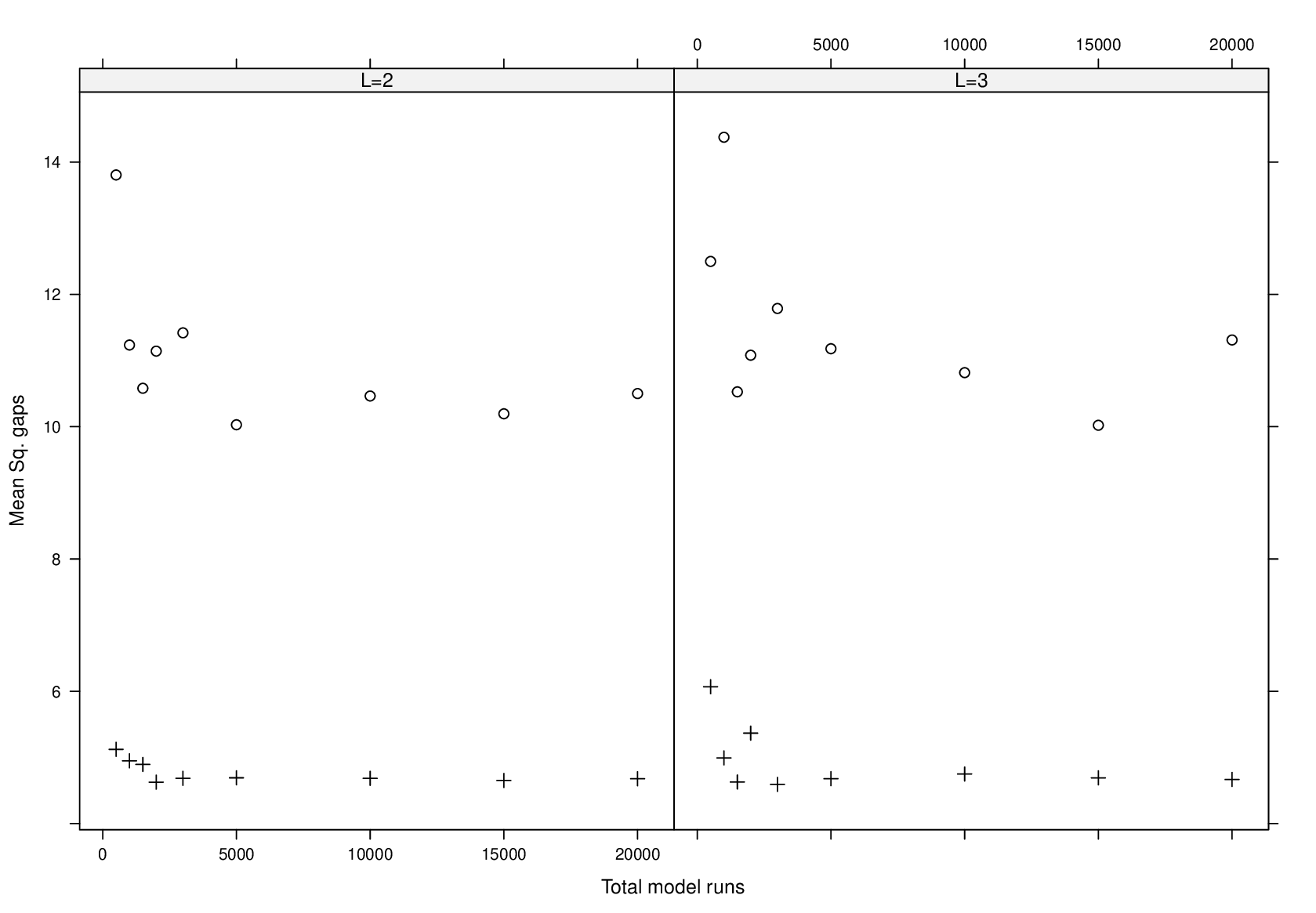}      
\end{center} 
\caption{Average of $d=3$ mean squared gaps using the Ishigami function ($\circ$ the direct estimator (\ref{eq:sigudfestub})   and $+$ for the plug-in estimator (\ref{eq:intsiguestub})).}         
 \label{fig:ishgap}          
\end{figure}

\begin{figure}[!hbp]         
\begin{center}
\includegraphics[height=15cm,width=9cm,angle=270]{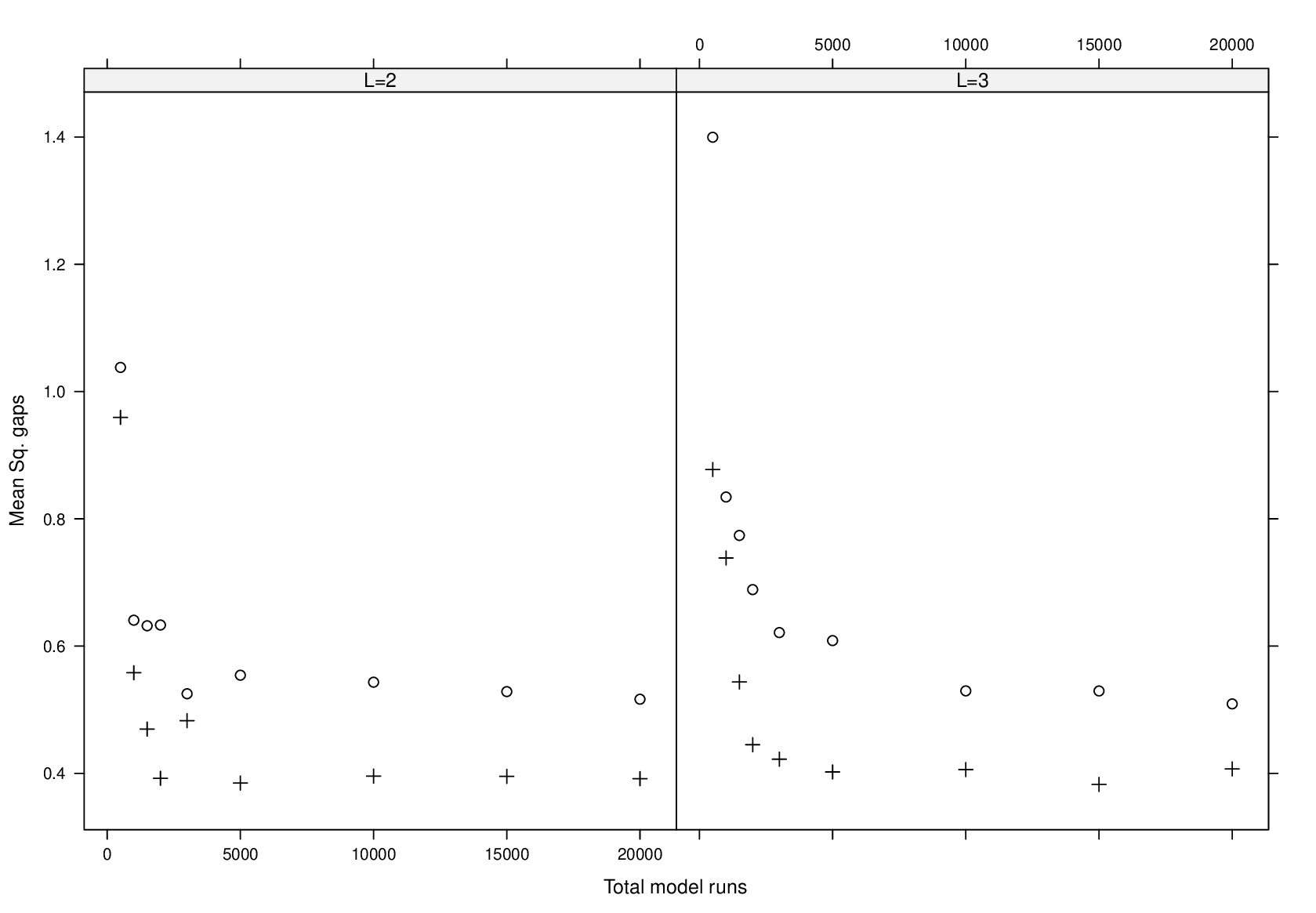}    
\end{center} 
\caption{Average of $d=10$ mean squared gaps using the g-function of type A ($\circ$ the direct estimator (\ref{eq:sigudfestub})   and $+$ for the plug-in estimator (\ref{eq:intsiguestub})).}      
 \label{fig:sobAgap} 
\end{figure}

\begin{figure}[!hbp]         
\begin{center}
\includegraphics[height=15cm,width=9cm,angle=270]{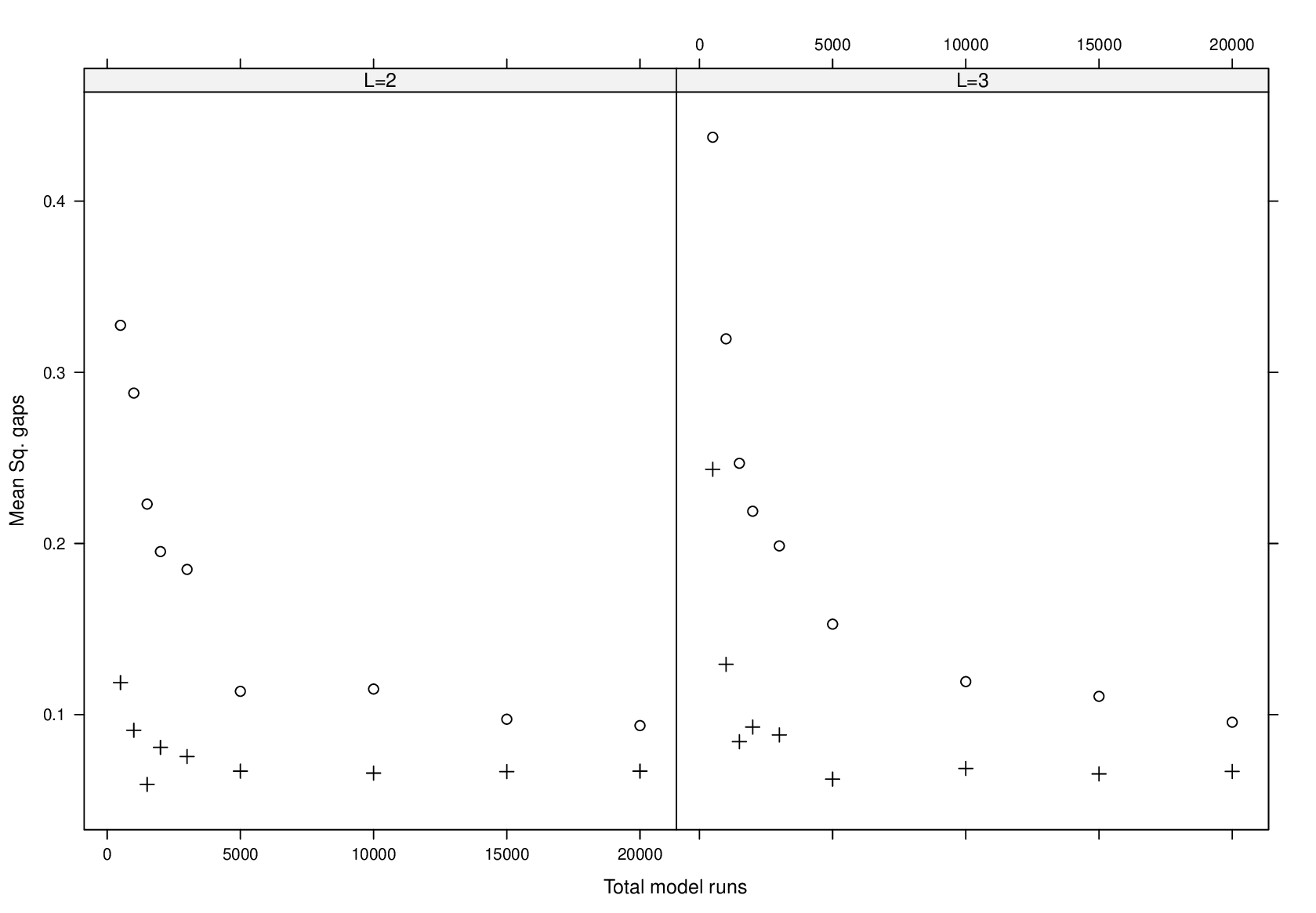}    
\end{center} 
\caption{Average of $d=10$ mean squared gaps using the g-function of type B ($\circ$ the direct estimator (\ref{eq:sigudfestub})   and $+$ for the plug-in estimator (\ref{eq:intsiguestub})).}      
 \label{fig:sobBgap}          
\end{figure}

\begin{figure}[!hbp]           
\begin{center}
\includegraphics[height=15cm,width=9cm,angle=270]{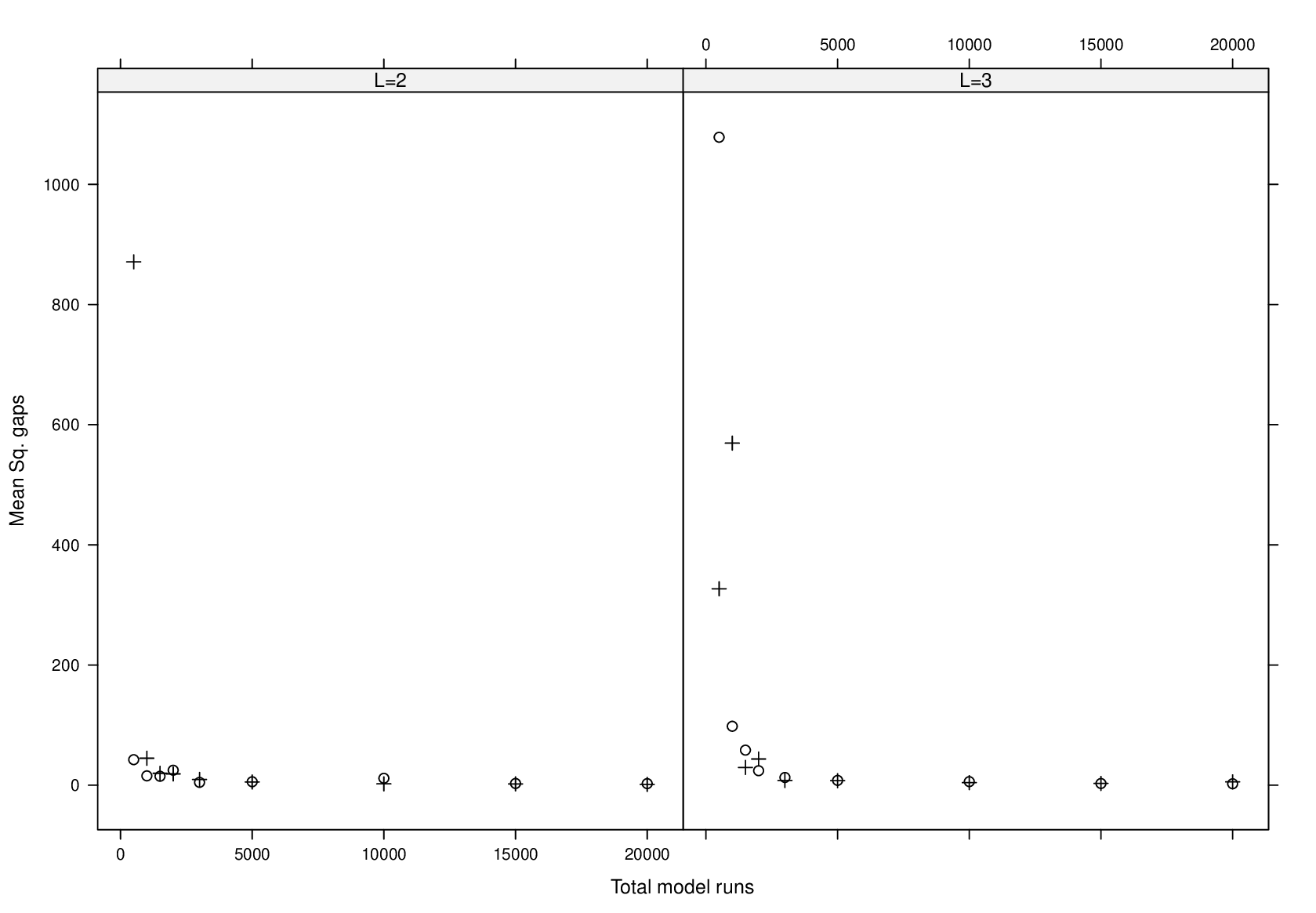}    
\end{center} 
\caption{Average of $d=10$ mean squared gaps using the g-function of type C ($\circ$ the direct estimator (\ref{eq:sigudfestub})   and $+$ for the plug-in estimator (\ref{eq:intsiguestub})).}         
 \label{fig:sobCgap}                  
\end{figure}

Likewise,  Figures \ref{fig:ishgap}-\ref{fig:sobCgap} show the mean squared gaps (differences) between the true total index and its estimated upper-bound for the Ishigami function, the g-functions of type A, type B and type C, respectively.      
   
It comes out that the plug-in estimators outperform  the direct ones . Also, increasing the values of $L$ gives the same results. Moreover,  the direct estimators associated with $L=1$ fail to provide accuracy estimates (we do not report such results here). In contrary, the plug-in estimates using $L=1$ are reported in Table~\ref{tab:ish} for the Ishigami function and in Table~\ref{tab:sobol} for the three types of the g-function. Such results
suggest considering $L=1$ or $L=2$ with $\beta_1=-1, \beta_2=1$ for plug-estimators when the total budget of model runs is small. For a larger budget of model runs, the direct estimators  associated with $L=2$ and $\beta_1=-1, \beta_2=1$ can be considered as well in practice.        
        
\begin{table}[htbp]  
\begin{center}   
\begin{tabular}{lcccc}   
\hline      
\hline    
 & X1 & X2 & X3 \\ 
  \hline
$S_j$ & 0.249 & 0.318 & -0.006 \\ 
$U\!B_j$ & 1.420 & 4.872 & 0.711 \\     
   \hline    \hline
\end{tabular}    
\end{center}   
\caption{Average of $30$ estimates of the main indices and upper-bounds of total indices for the Ishigami function using the plug-in estimators, $L=1$ and $2000$ model runs.}
\label{tab:ish}       
\end{table}      
    
\begin{table}[htbp]  
\begin{center}   
\begin{tabular}{lcccccccccc}   
\hline      
\hline          
 & X1 & X2 & X3 & X4 & X5 & X6 & X7 & X8 & X9 & X10 \\ 
  \hline
	 &  \multicolumn{10}{c}{Type A} \\ 
	\hline
$S_j$ & 0.330 & 0.324 & 0.006 & 0.005 & 0.006 & 0.006 & 0.005 & 0.005 & 0.006 & 0.005 \\ 
$U\!B_j$ & 2.022 & 2.005 & 0.045 & 0.046 & 0.047 & 0.047 & 0.046 & 0.046 & 0.046 & 0.047 \\ 
	  \hline
			 & \multicolumn{10}{c}{Type B} \\
	\hline
$S_j$ & 0.085 & 0.085 & 0.085 & 0.085 & 0.085 & 0.085 & 0.085 & 0.085 & 0.085 & 0.085 \\ 
$U\!B_j$ &  0.362 & 0.363 & 0.363 & 0.362 & 0.363 & 0.362 & 0.363 & 0.362 & 0.363 & 0.363 \\  
	  \hline
			 & \multicolumn{10}{c}{Type C} \\ 
	\hline
$S_j$ & 0.028 & 0.028 & 0.032 & 0.032 & 0.035 & 0.041 & 0.031 & 0.030 & 0.036 & 0.034 \\ 
$U\!B_j$ & 2.033 & 1.301 & 1.825 & 1.605 & 1.634 & 1.641 & 2.216 & 1.526 & 1.793 & 1.503 \\ 
   \hline  
	  \hline   
\end{tabular}
\end{center}   
\caption{Average of $30$ estimates of the main indices and upper-bounds of total indices for the g-functions using the plug-in estimators, $L=1$ and $2000$ model runs.}    
\label{tab:sobol}
\end{table}  
                                     
\subsection{Emulations of the g-function of type B}
\begin{figure}[!hbp]                 
\begin{center}
\includegraphics[height=15cm,width=10cm,angle=270]{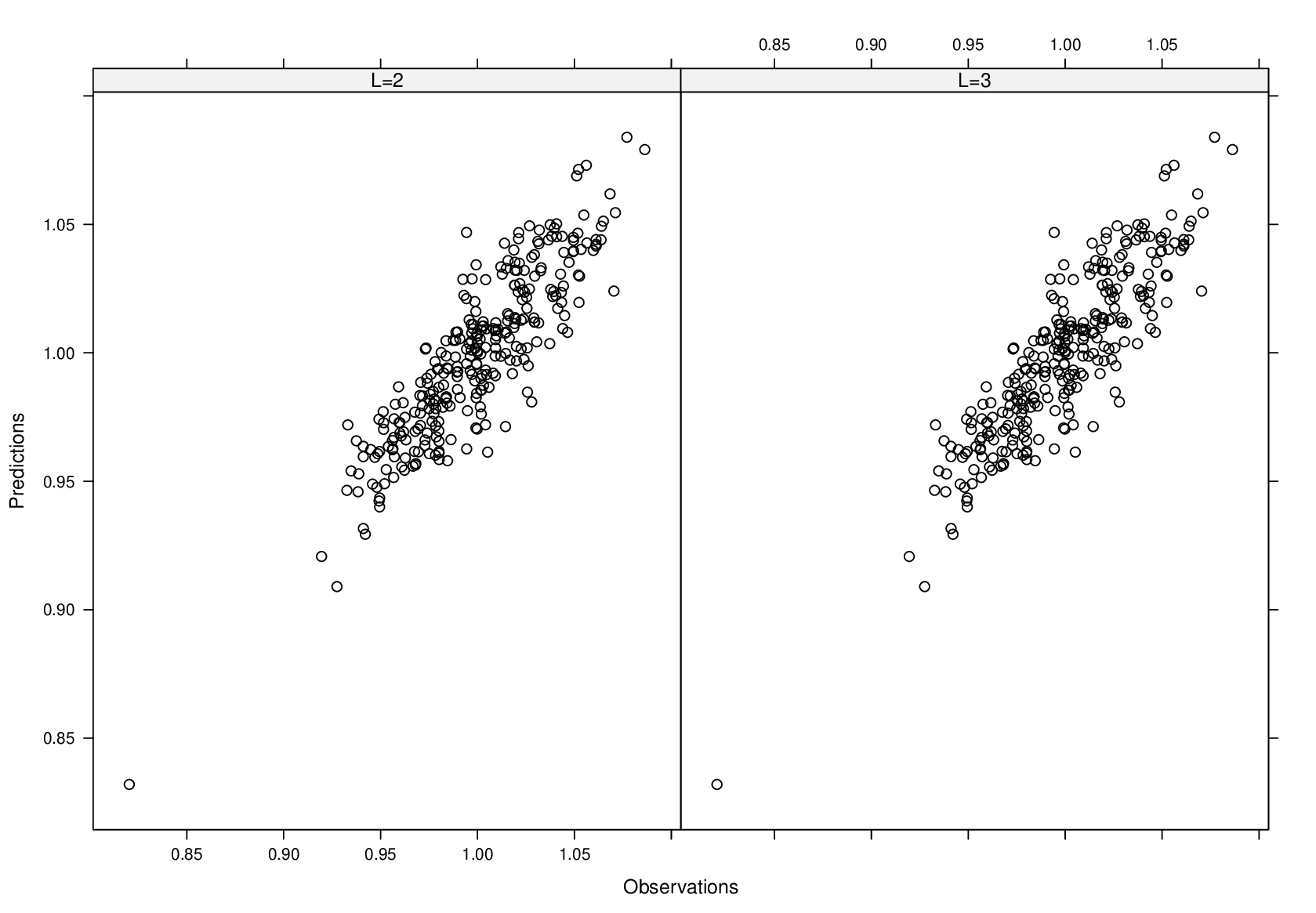}    
\end{center} 
\caption{Predictions of g-function of type B using the emulator $\widehat{\M_{s=1}}$ versus observations.}        
 \label{fig:emul}                
\end{figure}

Based on the results obtained in Section \ref{sec:indices} (see Table \ref{tab:sobol}), all the inputs are important in the case of the g-function of type B, meaning that the dimension reduction is not possible. Also,  the estimated upper-bounds suggest weak interactions among inputs. As expected, our estimated results confirm that the g-function of type B belongs to $\mathcal{A}_1$. Thus, an emulator of $\M$ based only on the first-order derivatives is sufficient. Using this information, we have derived the emulator of that function (i.e., $\widehat{\M_{s=1}}(\bo{x})$) under the assumptions made in Corollary \ref{coro:emu} ($r^*=0, L\geq 2$) and using $G_j=F_j$ with $j=1, \ldots, d$. For a given $L$, we used $300$ model runs to build the emulator, and Figure \ref{fig:emul} depicts the approximations of that function (called predictions) at the sample points involved in the construction of the emulator. Note that the evaluations of $\M$ at such sample points (called observations)  are not directly used in the construction of such an emulator. It comes out from  Figure \ref{fig:emul} that $\widehat{\M_{s=1}}$ provides predictions that are in line with the observations, showing the accuracy of our emulator.  
                   
\section{Conclusion} \label{sec:con}
In this paper, we have firstly provided i) stochastic expressions of cross-partial derivatives of any order, followed by their biases, and ii) estimators of such expressions. Our estimators of the $|u|$-th cross-partial derivatives ($\forall\, u \subseteq \{1, \ldots, d\}$) reach the parametric rates of convergence (i.e., $\mathsf{O}(N^{-1} d^{|u|})$) by means of a set of $L \geq |u|+1$ constraints for the H\"older  space of $\alpha$-smooth functions $\mathcal{H}_\alpha$ with $\alpha > |u|$. Moreover, we have shown that the upper-bounds of the biases of such estimators do not suffer from the curse of dimensionality. Secondly, the proposed surrogates of cross-partial derivatives are used for deriving  i) new derivative-based emulators of simulators or surrogates of models, even when a large number of model inputs contribute to the model outputs, and ii) new repressions of the main sensitivity indices and the upper-bounds of the total sensitivity indices. \\                
      
Numerical simulations have confirmed the accuracy of our approaches for not only screening the input variables, but also for identifying the class of functions that contains our simulator of interest, such as the class of functions with important or no interaction among inputs. This relevant information allows for designing and building the appropriate emulators of functions. In the case of the g-function of type B, our emulator of this function (based only on the first-order derivatives) has provided approximations or predictions that are in line with the observations. \\            
        
For functions with important interactions or equivalently for higher-order cross-partial derivatives, further numerical schemes are necessary to increase the numerical accuracy of the computations of such derivatives and predictions. Such perspectives are going to be investigated in next future as well as the computations of the total sensitivity indices using the proposed surrogates of derivatives.  Moreover, there is a need of a theoretical investigation so as to expect deriving the parametric rates of convergence of the above estimators that do not suffer from the course of dimensionality. Working in $\mathbb{C}$ rather than in $\R$ may be helpful.    
              
\begin{appendices}       
\section{Proof of Theorem \ref{theo:parderord}} \label{app:theo:parderord}
 Firstly, as $\Vec{\boldsymbol{\imath}} :=(i_1, \ldots, i_d)  \in \N^d$, denote $||\Vec{\boldsymbol{\imath}}||_1 = i_1 +\ldots+ i_{d}$ and $\Vec{\boldsymbol{u}} := \left(\indic_{u}(1), \ldots, \indic_{u}(d) \right)$. The Taylor expansion of  $\M\left(\bo{x} + \beta_\ell \boldsymbol{\hh}\bo{V} \right)$ about $\bo{x}$ of order $\alpha$ is given by   
\begin{eqnarray}                   
\M\left(\bo{x} + \beta_\ell \boldsymbol{\hh}\bo{V} \right)  &=&   \sum_{p=0}^{\alpha}  \sum_{||\Vec{\boldsymbol{\imath}}||_1= p}  \frac{\mathcal{D}^{(\Vec{\boldsymbol{\imath}})}\M(\bo{x}) }{\Vec{\boldsymbol{\imath}}!} \beta_\ell ^p  \left( \boldsymbol{\hh}\bo{V} \right)^{\Vec{\boldsymbol{\imath}}} +\mathcal{O}\left(||\beta_\ell \boldsymbol{\hh}\bo{V} ||_1^{\alpha+1} \right) \, . \nonumber                   
\end{eqnarray}          
Multiplying such an expansion by the constant $C_{\ell}^{(|u|)}$, and taking the sum over  $\ell=1, \ldots, L$, the expectation $E := \sum_{\ell=1}^L C_{\ell}^{(|u|)}\esp\left[ \M\left(\bo{x} + \beta_\ell \boldsymbol{\hh}\bo{V} \right) \prod_{k \in u} \frac{ V_k}{(\hh_k \sigma^2)} \right]$ becomes       
\begin{eqnarray}             
E  &=&  \sum_{p\geq 0}  \sum_{||\Vec{\boldsymbol{\imath}}||_1= p}  \frac{\mathcal{D}^{(\Vec{\boldsymbol{\imath}})}\M(\bo{x})}{\Vec{\boldsymbol{\imath}}!}	\left( \sum_{\ell} C_{\ell}^{(|u|)}  \beta_{\ell}^{p} \right)  \esp \left[ \frac{\left(\bo{V} \right)^{\Vec{\boldsymbol{\imath}} + \Vec{\boldsymbol{u}}}
\left( \boldsymbol{\hh} \right)^{\Vec{\boldsymbol{\imath}} -\Vec{\boldsymbol{u}}}}{\sigma^{2|u|}}	\right]  \, .  \nonumber   
\end{eqnarray}           
We can see that $\esp \left[ \left(\bo{V} \right)^{\Vec{\boldsymbol{\imath}} + \Vec{\boldsymbol{u}}}
\left( \boldsymbol{\hh} \right)^{\Vec{\boldsymbol{\imath}} -\Vec{\boldsymbol{u}}}	\right] \neq 0$ iff
$ \Vec{\boldsymbol{\imath}} + \Vec{\boldsymbol{u}} = 2\Vec{\bo{q}},\;  \forall \, \Vec{\bo{q}} \in \N^d$. 
Equation $\Vec{\boldsymbol{\imath}} + \Vec{\boldsymbol{u}} = 2\Vec{\bo{q}}$ implies  
$i_k =2q_k \geq 0$ if $k\notin u$ and $i_k=2q_k-1\geq 0$ otherwise. Thus, using $i_k =2q_k +1$ when $k \in u$ is much convenient, and it leads to $\Vec{\boldsymbol{\imath}} = 2\Vec{\bo{q}} + \Vec{\boldsymbol{u}}$, $\forall \, \Vec{\bo{q}} \in \N^d $, which also implies that $||  \Vec{\boldsymbol{\imath}}||_1 \geq ||\Vec{\boldsymbol{u}}||_1$.
 We then obtain $\mathcal{D}^{|u|}\M$ when $||\Vec{\bo{q}}||_1 =0$ or $\Vec{\boldsymbol{\imath}} = \Vec{\boldsymbol{u}}$,  and the fact that   
$  
\esp \left[ \left(\bo{V} \right)^{2 \Vec{\boldsymbol{u}}}	\right] = \esp \left[\prod_{k \in u} V_k^2	\right] =  \sigma^{2 |u|}        
$  by independence. We can then write            
$$
E = \mathcal{D}^{|u|}\M(\bo{x}) \left( \sum_{\ell} C_{\ell}^{(|u|)}  \beta_{\ell}^{|u|} \right)  +  \sum_{\substack{r\geq 1  \\||\Vec{\bo{q}}||_1= r}}  \frac{\mathcal{D}^{(2\Vec{\bo{q}} + \Vec{\boldsymbol{u}})}\M(\bo{x})}{(2\Vec{\bo{q}} + \Vec{\boldsymbol{u}})!} \left( \sum_{\ell} C_{\ell}^{(|u|)}  \beta_{\ell}^{r+|u|} \right) \esp \left[ \frac{\left(\bo{V} \right)^{2(\Vec{\boldsymbol{q}} + \Vec{\boldsymbol{u}})}
\left( \boldsymbol{\hh} \right)^{\Vec{2\boldsymbol{q}}}}{\sigma^{2|u|}}	\right]  \, , 
 $$        
using the change of variable $r=p-||\Vec{\boldsymbol{u}}||_1   $. At this point, setting $L=1, \, \beta_\ell=1$ and $C_{\ell}^{(|u|)}=1$ results in the  approximation of $\mathcal{D}^{|u|}\M(\bo{x})$ of order $\mathcal{O}(\norme{\hh}^{2})$.\\ 
Secondly, for $L>1$ the constraints $\sum_{\ell=1}^L C_{\ell}^{(|u|)}  \beta_{\ell}^{r+|u|} =\delta_{0, r}$ $r=0, 2, \ldots, 2(L-1)$ allow to eliminate some higher-order terms so as to reach the order $\mathcal{O}\left( \norme{\bo{\hh}}^{2L} \right)$. One can also use $\sum_{\ell=1}^L C_{\ell}^{(|u|)}  \beta_{\ell}^{r+|u|} =\delta_{0, r}$ $r=-|u|, \ldots, -|u| +L-2, 0$ to increase the accuracy of approximations, but keeping the order $\mathcal{O}\left( \norme{\bo{\hh}}^{2} \right)$ when $-|u| +L-2 < 0$. 
   
\section{Proof of Corollary \ref{coro:parderord0}} \label{app:coro:parderord0}
\noindent  
\noindent  
Let  $\bo{q} =(q_1, \ldots, q_d) \in \N^d$, $\Vec{\bo{u}} := \left(\indic_{u}(1), \ldots, \indic_{u}(d) \right) \in \N^d$, and consider  the set 
$ 
\boldsymbol{\alpha} := \left\{ 2\Vec{\bo{q}} + \Vec{\bo{u}} \; : \;  || \Vec{\bo{q}} ||_1 = L \right\}    
$.  
As $\M \in \mathcal{H}_{|u|+2L}$, the expansion of $\M\left(\bo{x} +  \boldsymbol{\hh}\bo{V} \right)$ gives       
$$   
\M(\bo{x}+ \beta_\ell  \boldsymbol{\hh}\bo{V}) =  \sum_{||\Vec{\boldsymbol{\imath}}||_1=0}^{|u|+2L-1} 
\mathcal{D}^{(\Vec{\boldsymbol{\imath}})}\M(\bo{x}) \beta_\ell^{||\Vec{\boldsymbol{\imath}}||_1}  \frac{(\boldsymbol{\hh}\bo{V})^{\Vec{\boldsymbol{\imath}}}}{\Vec{\boldsymbol{\imath}} !}   +
 \sum_{\substack{||\Vec{\boldsymbol{\imath}}||_1 =|u|+2L \\ \Vec{\boldsymbol{\imath}} \notin \boldsymbol{\alpha}}} \mathcal{D}^{(\Vec{\boldsymbol{\imath}})}\M(\bo{x}) \beta_\ell^{|u|+2L}  \frac{(\boldsymbol{\hh}\bo{V})^{\Vec{\boldsymbol{\imath}}}}{\Vec{\boldsymbol{\imath}} !}
+ R_{|u|}\left(\boldsymbol{\hh}, \beta_\ell, \bo{V} \right) \, ,   
$$      
with the  remainder term        
$ 
 R_{|u|}\left(\boldsymbol{\hh}, \beta_\ell, \bo{V} \right) := \beta_\ell^{|u|+2L}  \sum_{\substack{||\Vec{\boldsymbol{\imath}}||_1 =|u|+2L \\ \Vec{\boldsymbol{\imath}} \in \boldsymbol{\alpha}}} 
\mathcal{D}^{(\Vec{\boldsymbol{\imath}})}\M(\bo{x}+ \beta_\ell  \boldsymbol{\hh}\bo{V})  \frac{(\boldsymbol{\hh}\bo{V})^{\Vec{\boldsymbol{\imath}}}}{\Vec{\boldsymbol{\imath}} !} $ 
Thus,
$ 
 R_{|u|}\left(\boldsymbol{\hh}, \beta_\ell, \bo{V} \right) := \beta_\ell^{|u|+2L}  \sum_{\substack{||\Vec{\boldsymbol{q}}||_1 =L}} 
\frac{\mathcal{D}^{(2\Vec{\boldsymbol{q}} +\Vec{\boldsymbol{u}})}\M(\bo{x} + \beta_\ell  \boldsymbol{\hh}\bo{V})}{(2\Vec{\boldsymbol{q}}+ \Vec{\boldsymbol{u}}) !}(\boldsymbol{\hh}\bo{V})^{2\Vec{\boldsymbol{q}} +\Vec{\boldsymbol{u}}}$ and       
\begin{eqnarray}
 R_{|u|}\left(\boldsymbol{\hh}, \beta_\ell, \bo{V} \right) &=& \beta_\ell^{|u|+2L}  \sum_{\substack{||\Vec{\boldsymbol{q}}||_1 =L}} 
\frac{\mathcal{D}^{(2\Vec{\boldsymbol{q}} +\Vec{\boldsymbol{u}})}\M(\bo{x} + \beta_\ell  \boldsymbol{\hh}\bo{V})}{(2\Vec{\boldsymbol{q}}+ \Vec{\boldsymbol{u}}) !}(\boldsymbol{\hh}\bo{V})^{2\Vec{\boldsymbol{q}} +\Vec{\boldsymbol{u}}}  \nonumber \\
&=& \beta_\ell^{|u|+2L} (\boldsymbol{\hh}\bo{V})^{\Vec{\boldsymbol{u}}}   \sum_{\substack{||\Vec{\boldsymbol{q}}||_1 =L}} 
\frac{\mathcal{D}^{(2\Vec{\boldsymbol{q}} +\Vec{\boldsymbol{u}})}\M(\bo{x} + \beta_\ell  \boldsymbol{\hh}\bo{V})}{(2\Vec{\boldsymbol{q}}+ \Vec{\boldsymbol{u}}) !}(\boldsymbol{\hh}^2\bo{V}^2)^{\Vec{\boldsymbol{q}}}  \nonumber \, . 
\end{eqnarray}
Using $ E := \sum_{\ell=1}^{L=2} C_{\ell}^{(|u|)}\esp\left[ \M\left(\bo{x} + \beta_\ell \boldsymbol{\hh}\bo{V} \right) \prod_{k \in u} \frac{ V_k}{\hh_k \sigma^2} \right]$, Theorem \ref{theo:parderord} implies that the absolute value of the bias $B := \left|E - \mathcal{D}^{|u|}\M(\bo{x}) \right| = \left| \sum_{\ell=1}^L C_\ell^{(|u|)} \esp\left[ R_{|u|}\left(\boldsymbol{\hh}, \beta_\ell, \bo{V} \right) \prod_{k \in u} \frac{ V_k}{\hh_k \sigma^2} \right] \right|$ is given by 
$$       
B \leq  \sum_{\ell=1}^L  \left| C_\ell^{(|u|)}  \beta_\ell^{|u|+2L} \right| M_{|u|+2L} 
\esp\left[ ||\boldsymbol{\hh}^2\bo{V}^2 ||_1^L \prod_{k \in u} \frac{ V_k^2}{\sigma^2}  \right] \, ,
$$ 
as 
$
\left| \sum_{\substack{||\Vec{\boldsymbol{q}}||_1 =L}} 
\frac{\mathcal{D}^{(2\Vec{\boldsymbol{q}} +\Vec{\boldsymbol{u}})}\M(\bo{x} + \beta_\ell  \boldsymbol{\hh}\bo{V})}{(2\Vec{\boldsymbol{q}}+ \Vec{\boldsymbol{u}}) !}(\boldsymbol{\hh}^2\bo{V}^2)^{\Vec{\boldsymbol{q}}} \right| \leq M_{|u|+2L}  ||\boldsymbol{\hh}^2\bo{V}^2 ||_1^L      
$.   \\ 
Using $R_k=V_k/\sigma$, the results hold because $\esp\left[ ||\boldsymbol{\hh}^2\bo{V}^2 ||_1^L \prod_{k \in u} \frac{ V_k^2}{\sigma^2}  \right] = \sigma^{2L} \esp\left[||\boldsymbol{\hh}^2\bo{R}^2 ||_1^L \prod_{k \in u} R_k^2  \right]$. \\     
For $V_k \sim \mathcal{U}(-\xi, \xi)$ with $\xi>0$ and $k=1, \ldots,d$, we have  
$
\esp\left[ ||\boldsymbol{\hh}^2\bo{V}^2 ||_1^L \prod_{k \in u} \frac{ V_k^2}{\sigma^2}  \right] \leq
\xi^{2L}  ||\boldsymbol{\hh}^2||_1^L       
$.        
     
\section{Proof of Corollary \ref{coro:parderord}} \label{app:coro:parderord}
\noindent  
Using
$ 
\boldsymbol{\alpha} := \left\{ \Vec{\bo{q}} + \Vec{\bo{u}} \; : \;  || \Vec{\bo{q}} ||_1 = 1 \right\}    
$.  
As $\M \in \mathcal{H}_{|u|+1}^M$, we can write 
$$   
\M(\bo{x}+ \beta_\ell  \boldsymbol{\hh}\bo{V}) =  \sum_{||\Vec{\boldsymbol{\imath}}||_1=0}^{|u|} 
\mathcal{D}^{(\Vec{\boldsymbol{\imath}})}\M(\bo{x}) \beta_\ell^{||\Vec{\boldsymbol{\imath}}||_1}  \frac{(\boldsymbol{\hh}\bo{V})^{\Vec{\boldsymbol{\imath}}}}{\Vec{\boldsymbol{\imath}} !}   +
 \sum_{\substack{||\Vec{\boldsymbol{\imath}}||_1 =|u|+1 \\ \Vec{\boldsymbol{\imath}} \notin \boldsymbol{\alpha}}} 
\mathcal{D}^{(\Vec{\boldsymbol{\imath}})}\M(\bo{x}) \beta_\ell^{|u|+1}  \frac{(\boldsymbol{\hh}\bo{V})^{\Vec{\boldsymbol{\imath}}}}{\Vec{\boldsymbol{\imath}} !}
+ R_{|u|}\left(\boldsymbol{\hh}, \beta_\ell, \bo{V} \right) \, , 
$$      
with the  remainder term               
$ 
 R_{|u|}\left(\boldsymbol{\hh}, \beta_\ell, \bo{V} \right) := \beta_\ell^{|u|+1}  \sum_{\substack{||\Vec{\boldsymbol{\imath}}||_1 =|u|+1 \\ \Vec{\boldsymbol{\imath}} \in \boldsymbol{\alpha}}} 
\mathcal{D}^{(\Vec{\boldsymbol{\imath}})}\M(\bo{x}+ \beta_\ell  \boldsymbol{\hh}\bo{V})  \frac{(\boldsymbol{\hh}\bo{V})^{\Vec{\boldsymbol{\imath}}}}{\Vec{\boldsymbol{\imath}} !} 
$.   
Using $ E := \sum_{\ell=1}^{L=2} C_{\ell}^{(|u|)}\esp\left[ \M\left(\bo{x} + \beta_\ell \boldsymbol{\hh}\bo{V} \right) \prod_{k \in u} \frac{ V_k}{(\hh_k \sigma^2)} \right]$ and Theorem \ref{theo:parderord}, the results hold by analogy to the proof of Corollary \ref{coro:parderord0}. Indeed,  if $R_k := V_k/\sigma$, then   
\begin{eqnarray}   
B & \leq &     M_{|u|+1}     \esp\left[ || \boldsymbol{\hh} \bo{V} ||_1 \prod_{k \in u} \frac{ V_k^2}{\sigma^2}   \right] \Gamma_{|u|+1}  
 =   \sigma    M_{|u|+1} \, \esp\left[ || \boldsymbol{\hh} \bo{R} ||_1 \prod_{k \in u} R_k^2 \right] \Gamma_{|u|+1}   \nonumber \\ 
& \leq & \sigma  M_{|u|+1} \norme{\boldsymbol{\hh}}\, \esp\left[ \norme{\bo{R}} \prod_{k \in u} R_k^2 \right] \Gamma_{|u|+1}     \nonumber  \, .                    
\end{eqnarray}             
For $V_k \sim \mathcal{U}(-\xi, \xi)$ with $\xi>0$ and $k=1, \ldots,d$, we have 
$$  
 B \leq    M_{|u|+1}   
   \esp\left[ ||\boldsymbol{\hh} \bo{V}||_1 \prod_{k \in u} \frac{V_k^2}{\sigma^{2}}  \right] \Gamma_{|u|+1}   
\leq M_{|u|+1} \xi\, ||\boldsymbol{\hh}||_1 \esp\left[ \prod_{k \in u} \frac{ V_k^2}{\sigma^{2}}  \right] \Gamma_{|u|+1}  =  M_{|u|+1} ||\boldsymbol{\hh}||_1 \xi \Gamma_{|u|+1} \, .                    
$$          
                                      
\section{Proof of Theorem \ref{theo:mseall}} \label{app:theo:mseall}
As $\M \in \mathcal{H}_{|u|+1}$ implies that  $\M \in \mathcal{H}_{r^*+1}$ with $r^*\leq |u|-1$, we have 
$$
\left| \M(\bo{x}+ \beta_\ell  \boldsymbol{\hh}\bo{V}) -  \sum_{||\Vec{\boldsymbol{\imath}}||_1=0}^{r^*} 
\mathcal{D}^{(\Vec{\boldsymbol{\imath}})}\M(\bo{x}) \beta_\ell^{||\Vec{\boldsymbol{\imath}}||_1}  \frac{(\boldsymbol{\hh}\bo{V})^{\Vec{\boldsymbol{\imath}}}}{\Vec{\boldsymbol{\imath}} !} \right| \leq M_{r^*+1}\norme{\beta_\ell  \boldsymbol{\hh}\bo{V}}^{r^*+1} \, ,
$$    
Using the fact that $\sum_{\ell=1}^{L} C_\ell^{(|u|)} \beta_\ell^r =0$ for $r=0, 1, \ldots, r^*$, we can write   
$$ 
\sum_{\ell=1}^{L} C_\ell^{(|u|)} \M(\bo{x}+ \beta_\ell  \boldsymbol{\hh}\bo{V}) 
= \sum_{\ell=1}^{L} C_\ell^{(|u|)} \left[  \M(\bo{x}+ \beta_\ell  \boldsymbol{\hh}\bo{V}) -  \sum_{||\Vec{\boldsymbol{\imath}}||_1=0}^{r^*}    
\mathcal{D}^{(\Vec{\boldsymbol{\imath}})}\M(\bo{x}) \beta_\ell^{||\Vec{\boldsymbol{\imath}}||_1}  \frac{(\boldsymbol{\hh}\bo{V})^{\Vec{\boldsymbol{\imath}}}}{\Vec{\boldsymbol{\imath}} !} \right] \, , 
$$
which leads to      
$
 \left| \sum_{\ell=1}^{L}  C_\ell^{(|u|)} \M(\bo{x}+ \beta_\ell  \boldsymbol{\hh}\bo{V})  \right|  \leq   \sum_{\ell=1}^{L}  \left| C_\ell^{(|u|)} \beta_\ell^{r^*+1} \right| M_{r^*+1}\norme{\boldsymbol{\hh}\bo{V}}^{r^*+1}  
$. \\   
By taking the variance of the proposed estimator, we have    
\begin{eqnarray} 
\var \left[\widehat{\mathcal{D}^{|u|}\M}(\bo{x}) \right] & := &
\frac{1}{N} \var \left[  \left\{ \sum_{\ell=1}^{L} C_\ell^{(|u|)} \M(\bo{x}+ \beta_\ell  \boldsymbol{\hh}\bo{V}) \right\} \prod_{k \in u} \frac{ V_{k}}{(\hh_k \sigma^2)} \right] \nonumber \\
&\leq & \frac{1}{N} \esp \left[ \left\{ \sum_{\ell=1}^{L} C_\ell^{(|u|)} \M(\bo{x}+ \beta_\ell  \boldsymbol{\hh}\bo{V})  \right\}^2 \prod_{k \in u} \frac{ V_{k}^2}{(\hh_k^2 \sigma^4)} \right] \nonumber \\
 &\leq &  \frac{M_{r^*+1}^2 \left(\sum_{\ell=1}^{L} \left| C_\ell^{(|u|)}  \beta_\ell^{r^*+1} \right|\right)^2}{N\prod_{k \in u} \hh_k^2}     
\esp \left[\frac{\norme{\boldsymbol{\hh}\bo{V} }^{2(r^*+1)}}{\sigma^{2 |u|} } \prod_{k \in u} \frac{ V_{k}^2}{\sigma^2}   \right]  \nonumber \\            
&\leq&   \frac{M_{r^*+1}^2 \left(\sum_{\ell=1}^{L}  \left|  C_\ell^{(|u|)} \beta_\ell^{r^*+1} \right|\right)^2}{N \sigma^{2(|u|-r^*-1)}\prod_{k \in u} \hh_k^2}   \norme{\boldsymbol{\hh}^2}^{r^*+1} \esp \left[\norme{\bo{R}^2}^{r^*+1} \prod_{k \in u} R_{k}^2 \right] \nonumber \, ,              
\end{eqnarray}          
where $R_k=V_k/\sigma$.  \\               
 If $V_k \sim \mathcal{U}(-\xi, \xi)$,   
$\var \left[\widehat{\mathcal{D}^{|u|}\M}(\bo{x}) \right] \leq \frac{3^{|u|}M_{r^*+1}^2 \left(\sum_{\ell=1}^{L} \left| C_\ell^{(|u|)} \beta_\ell^{r^*+1} \right|\right)^2}{N \xi^{2(|u|-r^*-1)}\prod_{k \in u} \hh_k^2}  \norme{\boldsymbol{\hh}}^{2(r^*+1)} $.\\    
The results hold using Corollary \ref{coro:parderord} and the fact that $MSE= B^2 + \var \left[\widehat{\mathcal{D}^{|u|}\M}(\bo{x}) \right]$.        
  
\section{Proof of Corollary \ref{coro.optrate}} \label{app:coro.optrate}
Let $\eta := |u|-r^*-1>0$, $F_0:= 3^{|u|}M_{r^*+1}^2 \Gamma_{r^*+1}^2  d^{r^*+1}$ and $F_1 :=M_{|u|+1}^2   \Gamma_{|u|+1}^2 d^2$. By minimizing $\xi^2  M_{|u|+1}^2   \Gamma_{|u|+1}^2 d^2\hh^2 + \frac{3^{|u|}M_{r^*+1}^2 \Gamma_{r^*+1}^2  d^{r^*+1}}{N \xi^{2(|u|-r^*-1)} \hh^{2(|u|-r^*-1)}}$, we obtain  
$$
\hh_{op} = \frac{1}{\xi} \left(\frac{\eta F_0}{F_1} \right)^{\frac{1}{2\eta+2}} N^{-\frac{1}{2\eta+2}}; \qquad
\frac{F_0}{N \xi^{2\eta} \hh^{2\eta} } =  \frac{F_0}{N N^{-\frac{2\eta}{2\eta+2}}}  \left(\frac{F_1}{\eta F_0} \right)^{\frac{2\eta}{2\eta+2}} =  \left(\frac{F_0}{N}\right)^{\frac{1}{\eta+1}}  \left(\frac{F_1}{\eta} \right)^{\frac{\eta}{\eta+1}} \,  ;       
$$          
and     
$$  
d^{\frac{r^*+1}{|u|-r^*}}  d^{\frac{2(|u|-r^*-1)}{|u|-r^*}} = d^{\frac{2|u| -r^*-1}{|u|-r^*}} = d^{1+ \frac{|u|-1}{|u|-r^*}}  \, .  
$$      
    
\section{Proof of Theorem \ref{theo:mseaalp}} \label{app:theo:mseaalp} 
The first two results hold by combining the biases obtained in Corollary \ref{coro:parderord0} and the upper-bounds of the variance provided in Theorem  \ref{theo:mseall}.   \\  
For the last result, let $\eta := |u|-r^*-1>0$, $F_0:= 3^{|u|}M_{r^*+1}^2 \Gamma_{r^*+1}^2  d^{r^*+1}$ and $F_1 :=M_{|u|+2L'}^2  \Gamma_{|u|+2L'}^2 d^{2L'}$. By minimizing the last upper-bound, we obtain    $\hh_{op} := \frac{1}{\xi} \left(\frac{\eta F_0}{2L'F_1} \right)^{\frac{1}{2\eta+4L'}} N^{-\frac{1}{2\eta+4l'}}$ and    
$$ 
\frac{F_0}{N \xi^{2\eta} \hh^{2\eta} } =  \frac{F_0}{N N^{-\frac{2\eta}{2\eta+4L'}}}  \left(\frac{2L' F_1}{\eta F_0} \right)^{\frac{2\eta}{2\eta+4L'}} =  \left(\frac{F_0}{N}\right)^{\frac{2L'}{\eta+2L'}}  \left(\frac{2L'F_1}{\eta} \right)^{\frac{\eta}{\eta+2L'}} \propto N^{-\frac{2L'}{|u|-r^*-1+2L'}} d^{\frac{2L'|u|}{|u|-r^*-1+2L'}}  \,  .          
$$              
              
\section{On Remark \ref{rem:bo}} \label{app:rem:bo}   
The variance is   
$
\var\left[\widehat{\mathcal{D}^{|u|}\M(\bo{x})} \right]  =
\frac{1}{N} \var \left[\sum_{\ell=1}^{L} C_{\ell}^{(|u|)} \M\left(\bo{x} + \beta_\ell \boldsymbol{\hh}\bo{V} \right) \prod_{k \in u} \frac{ V_{k}}{\hh_k \sigma^2} \right]  
$ 
and     
$$
\var\left[\widehat{\mathcal{D}^{|u|}\M(\bo{x})} \right] \leq \frac{1}{N} \esp \left[\left(\sum_{\ell=1}^{L} C_{\ell}^{(|u|)}\M\left(\bo{x} + \beta_\ell \boldsymbol{\hh}\bo{V} \right)\right)^2 \prod_{k \in u} \frac{V_{k}^2}{\hh_k^2 \sigma^4} \right]  
\leq  \frac{ \norminf{\M}^2 \sum_{\substack{\ell_1=1, \ell_2=1}}^{L} C_{\ell_1}^{(|u|)} C_{\ell_2}^{(|u|)}}{N \hh^{2|u|} \sigma^{2|u|}}\,  .         
$$       
     
\end{appendices}                                  
       

\begin{thebibliography}{10}
\expandafter\ifx\csname url\endcsname\relax
  \def\url#1{\texttt{#1}}\fi
\expandafter\ifx\csname urlprefix\endcsname\relax\def\urlprefix{URL }\fi
\expandafter\ifx\csname href\endcsname\relax
  \def\href#1#2{#2} \def\path#1{#1}\fi

\bibitem{robbins51}
H.~Robbins, S.~Monro, {A Stochastic Approximation Method}, The Annals of
  Mathematical Statistics 22~(3) (1951) 400 -- 407.

\bibitem{fabian71}
V.~Fabian, Stochastic approximation, in: Optimizing methods in statistics,
  Elsevier, 1971, pp. 439--470.

\bibitem{nemirovsky83}
A.~Nemirovsky, D.~Yudin, Problem Complexity and Method Efficiency in
  Optimization, Wiley \& Sons, New York, 1983.

\bibitem{polyak90}
B.~Polyak, A.~Tsybakov, Optimal accuracy orders of stochastic approximation
  algorithms, Probl. Peredachi Inf. (1990) 45--53.

\bibitem{cristea17}
M.~Cristea, On global implicit function theorem, Journal of Mathematical
  Analysis and Applications 456~(2) (2017) 1290--1302.

\bibitem{lamboni23axioms}
M.~Lamboni, Derivative formulas and gradient of functions with non-independent
  variables, Axioms 12~(9) (2023).
\newblock \href {https://doi.org/10.3390/axioms12090845}
  {\path{doi:10.3390/axioms12090845}}.

\bibitem{morris93}
T.~J.~M. Max D.~Morris, D.~Ylvisaker, Bayesian design and analysis of computer
  experiments: Use of derivatives in surface prediction, Technometrics 35~(3)
  (1993) 243--255.

\bibitem{solak02}
E.~Solak, R.~Murray-Smith, W.~Leithead, D.~Leith, C.~Rasmussen, Derivative
  observations in {G}aussian process models of dynamic systems, Advances in
  neural information processing systems 15 (2002).

\bibitem{hoeffding48a}
W.~Hoeffding, A class of statistics with asymptotically normal distribution,
  Annals of Mathematical Statistics 19 (1948) 293--325.

\bibitem{efron81}
B.~Efron, C.~Stein, The jacknife estimate of variance, The Annals of Statistics
  9 (1981) 586--596.

\bibitem{sobol93}
I.~M. Sobol, Sensitivity analysis for non-linear mathematical models,
  Mathematical Modelling and Computational Experiments 1 (1993) 407--414.

\bibitem{rabitz99}
H.~Rabitz, General foundations of high dimensional model representations,
  Journal of Mathematical Chemistry 25 (1999) 197--233.

\bibitem{saltelli00}
A.~Saltelli, K.~Chan, E.~Scott, Variance-Based Methods, Probability and
  Statistics, John Wiley and Sons, 2000.

\bibitem{lamboni22}
M.~Lamboni, Weak derivative-based expansion of functions: {ANOVA} and some
  inequalities, Mathematics and Computers in Simulation 194 (2022) 691--718.

\bibitem{sobol09}
I.~M. Sobol, S.~Kucherenko, Derivative based global sensitivity measures and
  the link with global sensitivity indices, Mathematics and Computers in
  Simulation 79 (2009) 3009--3017.

\bibitem{kucherenko09}
S.~Kucherenko, M.~Rodriguez-Fernandez, C.~Pantelides, N.~Shah, {Monte Carlo}
  evaluation of derivative-based global sensitivity measures, Reliability
  Engineering and System Safety 94 (2009) 1135--1148.

\bibitem{lamboni13}
M.~Lamboni, B.~Iooss, A.-L. Popelin, F.~Gamboa, Derivative-based global
  sensitivity measures: General links with {S}obol' indices and numerical
  tests, Mathematics and Computers in Simulation 87~(0) (2013) 45 -- 54.

\bibitem{roustant14}
O.~Roustant, J.~Fruth, B.~Iooss, S.~Kuhnt, Crossed-derivative based sensitivity
  measures for interaction screening, Mathematics and Computers in Simulation
  105 (2014) 105 -- 118.

\bibitem{lamboni19}
M.~Lamboni, Derivative-based generalized sensitivity indices and {S}obol'
  indices, Mathematics and Computers in Simulation 170 (2020) 236 -- 256.

\bibitem{lamboni21}
M.~Lamboni, S.~Kucherenko, Multivariate sensitivity analysis and
  derivative-based global sensitivity measures with dependent variables,
  Reliability Engineering \& System Safety 212 (2021) 107519.

\bibitem{lamboni23mcap}
M.~Lamboni, On exact distribution for multivariate weighted distributions and
  classification, Methodology and Computing in Applied Probability 25 (2023)
  1--41.

\bibitem{lamboni24uq}
M.~Lamboni, Measuring inputs-outputs association for time-dependent hazard
  models under safety objectives using kernels, International Journal for
  Uncertainty Quantification - (2024) 1--17.
\newblock \href
  {https://doi.org/10.1615/Int.J.UncertaintyQuantification.2024049119}
  {\path{doi:10.1615/Int.J.UncertaintyQuantification.2024049119}}.

\bibitem{lamboni24}
M.~Lamboni, Kernel-based measures of association between inputs and outputs
  using {ANOVA}, Sankhya A - (2024).
\newblock \href {https://doi.org/10.1007/s13171-024-00354-w}
  {\path{doi:10.1007/s13171-024-00354-w}}.

\bibitem{russi10}
T.~M. Russi, Uncertainty quantification with experimental data and complex
  system models, 2010.

\bibitem{constantine14}
P.~Constantine, E.~Dow, S.~Wang, Active subspace methods in theory and
  practice: Applications to kriging surfaces, SIAM Journal on Scientific
  Computing 36 (2014) 1500--1524.

\bibitem{zahm20}
O.~Zahm, P.~G. Constantine, C.~Prieur, Y.~M. Marzouk, Gradient-based dimension
  reduction of multivariate vector-valued functions, SIAM Journal on Scientific
  Computing 42~(1) (2020) A534--A558.

\bibitem{kubicek15}
M.~Kubicek, E.~Minisci, M.~Cisternino, High dimensional sensitivity analysis
  using surrogate modeling and high dimensional model representation,
  International Journal for Uncertainty Quantification 5~(5) (2015) 393--414.

\bibitem{kuo10}
F.~Kuo, I.~Sloan, G.~Wasilkowski, H.~Wo{\'z}niakowski, On decompositions of
  multivariate functions, Mathematics of computation 79~(270) (2010) 953--966.

\bibitem{bates80}
D.~Bates, D.~Watts, Relative curvature measures of nonlinearity, J. Royal
  Statistics Soc. series B 42 (1980) 1--25.

\bibitem{guidotti22}
E.~Guidotti, {calculus}: High-dimensional numerical and symbolic calculus in
  {R}, Journal of Statistical Software 104~(5) (2022) 1--37.

\bibitem{dimet86}
F.-X. Le~Dimet, O.~Talagrand, Variational algorithms for analysis and
  assimilation of meteorological observations: theoretical aspects, Tellus A:
  Dynamic Meteorology and Oceanography 38~(2) (1986) 97--110.

\bibitem{dimet97}
F.~X. Le~Dimet, H.~E. Ngodock, B.~Luong, J.~Verron, Sensitivity analysis in
  variational data assimilation, Journal-Meteorological Society of Japan 75
  (1997) 245--255.

\bibitem{cacuci05}
D.~G. Cacuci, Sensitivity and uncertainty analysis - Theory, Chapman \&
  Hall/CRC, 2005.

\bibitem{gunzburger03}
M.~D. Gunzburger, Perspectives in flow control and optimization, SIAM,
  Philadelphia, 2003.

\bibitem{borzi12}
A.~Borzi, V.~Schulz, Computational Optimization of Systems Governed by Partial
  Differential Equations, SIAM, Philadelphia, 2012.

\bibitem{ghanem17}
R.~Ghanem, D.~Higdon, H.~Owhadi, Handbook of Uncertainty Quantification,
  Springer International Publishing, 2017.

\bibitem{wang92}
Z.~Wang, I.~M. Navon, F.-X. Le~Dimet, X.~Zou, The second order adjoint
  analysis: theory and applications, Meteorology and atmospheric physics 50
  (1992) 3--20.

\bibitem{agarwal10}
A.~Agarwal, O.~Dekel, L.~Xiao, Optimal algorithms for online convex
  optimization with multi-point bandit feedback., in: Colt, Citeseer, 2010, pp.
  28--40.

\bibitem{bach16}
F.~Bach, V.~Perchet, Highly-smooth zero-th order online optimization, in:
  V.~Feldman, A.~Rakhlin, O.~Shamir (Eds.), 29th Annual Conference on Learning
  Theory, Vol.~49, 2016, pp. 257--283.

\bibitem{akhavan20}
A.~Akhavan, M.~Pontil, A.~B. Tsybakov, Exploiting higher order smoothness in
  derivative-free optimization and continuous bandits, NIPS '20, Curran
  Associates Inc., Red Hook, NY, USA, 2020.

\bibitem{lamboni24axioms}
M.~Lamboni, Optimal and efficient approximations of gradients of functions with
  nonindependent variables, Axioms 13~(7) (2024).
\newblock \href {https://doi.org/10.3390/axioms13070426}
  {\path{doi:10.3390/axioms13070426}}.

\bibitem{patelli10}
E.~Patelli, H.~Pradlwarter, {Monte Carlo} gradient estimation in high
  dimensions, International Journal for Numerical Methods in Engineering 81~(2)
  (2010) 172--188.

\bibitem{prashanth16}
L.~Prashanth, S.~Bhatnagar, M.~Fu, S.~Marcus, Adaptive system optimization
  using random directions stochastic approximation, IEEE Transactions on
  Automatic Control 62~(5) (2016) 2223--2238.

\bibitem{agarwal17}
N.~Agarwal, B.~Bullins, E.~Hazan, Second-order stochastic optimization for
  machine learning in linear time, J. Mach. Learn. Res. 18~(1) (2017)
  4148--4187.

\bibitem{zhu20}
J.~Zhu, L.~Wang, J.~C. Spall, Efficient implementation of second-order
  stochastic approximation algorithms in high-dimensional problems, IEEE
  Transactions on Neural Networks and Learning Systems 31~(8) (2020)
  3087--3099.

\bibitem{zhu22}
J.~Zhu, Hessian estimation via Stein's identity in black-box problems, in:
  J.~Bruna, J.~Hesthaven, L.~Zdeborova (Eds.), Proceedings of the 2nd
  Mathematical and Scientific Machine Learning Conference, Vol. 145 of
  Proceedings of Machine Learning Research, PMLR, 2022, pp. 1161--1178.

\bibitem{erdogdu15}
M.~A. Erdogdu, Newton-stein method: A second order method for glms via
  stein\textquotesingle s lemma, in: C.~Cortes, N.~Lawrence, D.~Lee,
  M.~Sugiyama, R.~Garnett (Eds.), Advances in Neural Information Processing
  Systems, Vol.~28, Curran Associates, Inc., 2015.

\bibitem{stein04}
C.~Stein, P.~Diaconis, S.~Holmes, G.~Reinert, Use of exchangeable pairs in the
  analysis of simulations, Lecture Notes-Monograph Series 46 (2004) 1--26.

\bibitem{zemanian87}
A.~Zemanian, Distribution Theory and Transform Analysis: An Introduction to
  Generalized Functions, with Applications, Dover Books on Advanced
  Mathematics, Dover Publications, 1987.

\bibitem{strichartz94}
R.~Strichartz, A Guide to Distribution Theory and Fourier Transforms, Studies
  in advanced mathematics, CRC Press, Boca, 1994.

\bibitem{rawashdeh19}
E.~Rawashdeh, A simple method for finding the inverse matrix of {V}andermonde
  matrix, Mathematiqki Vesnik 71 (2019) 207--213.

\bibitem{ahmed23}
A.~Arafat, M.~El-Mikkawy, A fast novel recursive algorithm for computing the
  inverse of a generalized {V}andermonde matrix, Axioms 12~(1) (2023).

\bibitem{morris91}
M.~Morris, Factorial sampling plans for preliminary computational experiments,
  Technometrics 33 (1991) 161--174.

\bibitem{roustant17}
O.~Roustant, F.~Barthe, B.~Iooss, Poincaré inequalities on intervals -
  application to sensitivity analysis, Electron. J. Statist. 11~(2) (2017)
  3081--3119.

\bibitem{lamboni16}
M.~Lamboni, Global sensitivity analysis: an efficient numerical method for
  approximating the total sensitivity index, International Journal for
  Uncertainty Quantification 6~(1) (2016) 1--17.

\bibitem{lamboni18a}
M.~Lamboni, Multivariate sensitivity analysis: Minimum variance unbiased
  estimators of the first-order and total-effect covariance matrices,
  Reliability Engineering \& System Safety 187 (2019) 67 -- 92.

\bibitem{lamboni18}
M.~Lamboni, Uncertainty quantification: a minimum variance unbiased (joint)
  estimator of the non-normalized {S}obol' indices, Statistical Papers (2018)
  --\href {https://doi.org/https://doi.org/10.1007/s00362-018-1010-4}
  {\path{doi:https://doi.org/10.1007/s00362-018-1010-4}}.

\bibitem{homma96}
T.~Homma, A.~Saltelli, Importance measures in global sensitivity analysis of
  nonlinear models, Reliability Engineering and System Safety 52 (1996) 1--17.

\bibitem{dutang13}
C.~Dutang, P.~Savicky, randtoolbox: Generating and Testing Random Numbers, {\sc
  R} package version 1.13 (2013).

\end{thebibliography}

\end{document}